# The TOFp/pVPD Time-Of-Flight System for STAR.

W.J. Llope, [a,*] F. Geurts, [a] J.W. Mitchell, [b] Z. Liu, [c] N. Adams, [a] G. Eppley, [a] D. Keane, [e] J. Li, [d] F. Liu, [c] L. Liu, [c] G.S. Mutchler, [a] T. Nussbaum, [a] B. Bonner, [a] P. Sappenfield, [a] B. Zhang, [d] W.-M. Zhang. [e]

[a] *T.W. Bonner Nuclear Laboratory, Rice University, Houston, TX 77005*

[b] *NASA–Goddard Space Flight Center, Code 661, Greenbelt, MD 20771*

[c] *HuaZhong Normal University, Wuhan, China*

[d] *Institute of High Energy Physics, Beijing, China*

[e] *Kent State University, Kent, OH 44242*

**Abstract**

A time-of-flight system was constructed for the STAR Experiment for the direct identification of hadrons produced in $^{197}$Au+$^{197}$Au collisions at RHIC. The system consists of two separate detector subsystems, one called the pVPD (the “start” detector) and the other called the TOFp tray (the “stop” detector). Each detector is based on conventional scintillator/phototube technology and includes custom high-performance front-end electronics and a common CAMAC-based digitization and read-out. The design of the system and its performance during the 2001 RHIC run will be described. The start resolution attained by the pVPD was 24 ps, implying a pVPD single-detector resolution of 58 ps. The total time resolution of the system averaged over all detector channels was 87 ps, allowing direct $\pi$/K/p discrimination for momenta up to $\sim$1.8 GeV/c, and direct ($\pi$+K)/p discrimination up to $\sim$3 GeV/c.

*Key words:* Particle identification; Time of Flight; STAR.

*PACS:* 06.60.Jn, 07.50.Ek, 29.40.Mc

* Corresponding author. Tel.: +1-713-348-4741, Fax.: +1-713-348-5215.
*E-mail address:* llope@physics.rice.edu (W.J. Llope).

## 1 Introduction

The Solenoidal Tracker (STAR) [1] at the Relativistic Heavy-Ion Collider (RHIC) investigates the behavior of strongly interacting matter at high density, and searches for the possible formation of the Quark-Gluon Plasma. The STAR design involves a large and azimuthally complete acceptance for charged hadrons, composed primarily of $\pi$ and K mesons and protons, produced in RHIC collisions. The heart of the STAR detector is a large Time Projection Chamber (TPC) [2] inside a 0.5T solenoidal magnetic field [3]. Charged hadron Particle Identification (PID) in STAR is possible via dE/dx in the TPC, and in a small acceptance, by a Ring Imaging Čerenkov (RICH) detector [4]. The ranges of particle momenta, p, over which these detectors provide PID information are disjoint. Kaons(protons) are identifiable via the TPC over the range $0.1\lesssim p\lesssim 0.7$ GeV/c ($0.1\lesssim p\lesssim 1.0$ GeV/c), and via the RICH over the range $1.0\lesssim p\lesssim 3.0$ GeV/c ($1.5\lesssim p\lesssim 5.0$ GeV/c), *i.e.* TPC PID exists at low momenta, RICH PID exists at high momenta, and a "PID blind" gap exists in between.

In RHIC $^{197}$Au+$^{197}$Au collisions and for particle transverse momenta below about 1-2 GeV/c, the particle production proceeds via soft, non-perturbative, mechanisms with the produced hadrons interacting and producing more hadrons during the expansion and cooling of the system. These interactions may generate collective behavior such as transverse radial flow and elliptic flow. A key question is the competition between this soft production and the fragmentation products from hard processes, *i.e.* collisions between partons, which are expected eventually to dominate the particle production at sufficiently high transverse momenta. Particles in different momentum intervals thus tend to have experienced different environments in different epochs during the evolution of the system, and also different degrees of final-state interactions with other particles. The momentum regions missing in the STAR PID acceptance is just in this interesting "transitional" region of the spectra. Mapping out the transition from soft to hard particle production is critical for a full understanding of the collision dynamics. Identified particle momentum spectra with a coverage that is continuous in momentum, and with the redundancy of multiple PID techniques in limited momentum ranges, are more valuable than those with holes in the PID acceptance in the key regions.

In order to close the gap in the STAR PID capabilities versus the particle momentum, and to measure directly-identified charged hadron spectra over wide momentum ranges, we added a small-acceptance Time-Of-Flight (TOF) system to the STAR detector. The system consists of two separate detector subsystems, one called the pVPD (the "start" detector) and the other called the TOFp (the "stop" detector). Each of these is based on conventional scintillator/phototube technology and includes custom high-performance front end electronics and common digitization in CAMAC. The inaugural experimen-tal run of the system was in the Fall of 2001, when colliding beams of 100 GeV/c/N $^{197}$Au and 100 GeV/c/N $^{197}$Au were available for the first time.

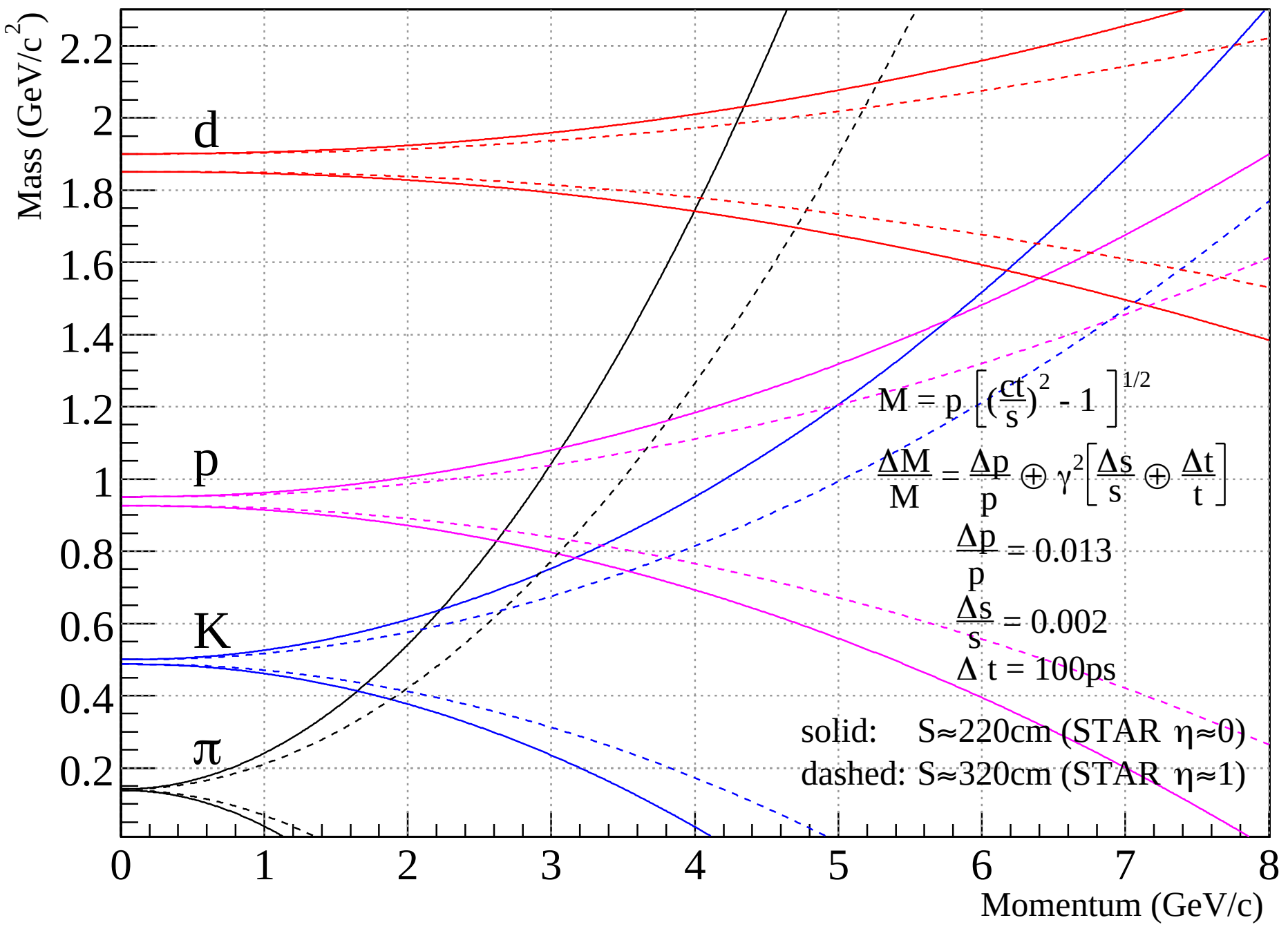

Fig. 1. The momentum dependence of the TOF mass resolution for the STAR apparatus.

In terms of its use for the particle tracking in a TOF analysis in STAR, the TPC [2] is extremely powerful detector. The TPC is a large cylinder with an outer(inner) diameter of 4(1) m which extends ±2.1 m on either side of the beam intersection region, or approximately two units of pseudorapidity from -1 to 1. Its track momentum resolution, Δp/p, is 1-2%, and its resolution on track total path lengths is <0.5cm. A TOF system in STAR at the cylindrical radius of the TPC with a total time interval resolution of 100 ps thus has the PID capabilities indicated in Figure 1. The figure shows the particle mass determined via TOF for each of four charged hadrons of interest - pions, Kaons, protons, and deuterons, as labelled, versus the particle momentum for the STAR geometry and for the momentum and path length resolutions of the STAR TPC. Plotted for each particle are a pair of lines indicating the dependence of M+ΔM (upper line in pair) and M−ΔM (lower line in pair) versus the momentum. The solid pairs of lines correspond to tracks in STAR near pseudorapidity $\eta$∼0 (the center of the TPC), while the dashed pairs of lines correspond to tracks near pseudorapidity $\eta$∼1, which have a ∼50% longer path length. The momentum at which an upper line for a particle touches the lower line for another particle is the maximum momentum for which "2$\sigma$" PID is possible via a STAR TOF system with a 100 ps resolution. From the figure, such a system should provide direct $\pi$/K/p identification up to momenta of ∼1.7(1.9) GeV/c, and direct p *vs.* ($\pi$+K) identification up to ∼2.6(3.1) GeV/c, for tracks near $\eta\sim0(1)$. Deuterons can be directly identified out to $\sim4(4.7)$ GeV/c. One notes from this figure that, with increasing momentum, pions, not Kaons, are the first particles to lead to a significant background in the proton identification. Similarly, pions, not protons, are also the first background to deuteron identification.

## 2 System Description

Described in this section is the hardware for the TOFp/pVPD Time Of Flight system. The detectors consist of two separate STAR subsystems. The electronic signals from these detectors define the time intervals of interest for particle Time of Flight measurements - the Pseudo Vertex Position Detector (pVPD) is the start detector and the Time-Of-Flight Patch (TOFp) is the stop detector. The arrangement of these systems with respect to the STAR TPC and the RHIC beam pipe is shown in Figure 2. The pVPD consists of two identical detector assemblies that are positioned very close to the beam pipe and outside the STAR magnet. The TOFp sits inside the STAR magnet immediately outside the TPC. The signals from these detectors are carried to electronics racks on the so-called South Platform next to STAR for digitization and interfacing with the STAR data stream.

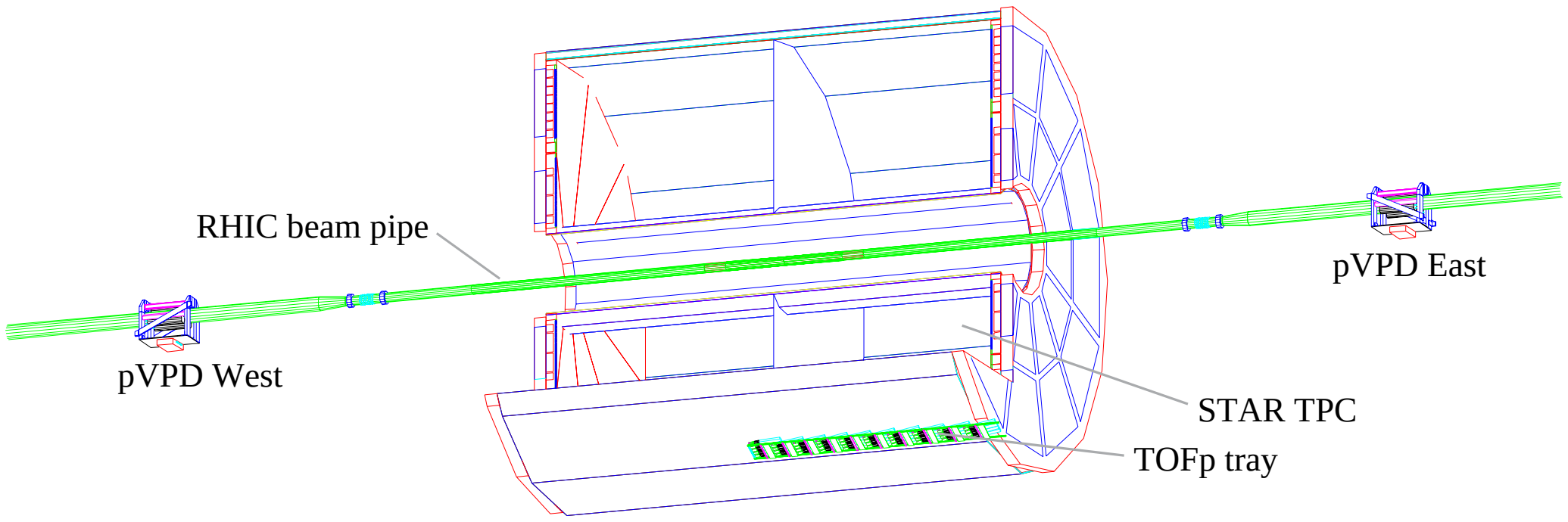

Fig. 2. A scale drawing of the locations of pVPD and TOFp detectors in relation to the STAR TPC and the RHIC beam pipe. For clarity, the TPC is cut away, while the STAR magnet and other subsystems are not drawn.

In RHIC full-energy Au+Au collisions, large numbers of very forward, very high energy, photons are produced which travel away from the collision vertex effectively as a prompt pulse. Measuring the times when these very forward particle pulses arrive at equally-spaced detectors on each side of STAR provides the location of the collision vertex along the beam pipe. The average of these two arrival times is the event start time, which with the TOFp stop times provides the time interval measurements of interest. To perform this function we implemented another new detector for STAR called the Pseudo Vertex Position Detector (pVPD). The design of this detector is based on the conventional technology of plastic scintillator read-out by photomultiplier tubes. There are three pVPD detector elements on each side of STAR at a distance of approximately 5 m from the intersection region.

The detector measuring the particles of interest - the stop detector - is the Time of Flight Patch (TOFp). The goal of the system is to provide direct particle identification for the formation of the rates and spectra of charged hadrons produced in RHIC in a small acceptance but over many events. The detector design is also based on the conventional technology. A total of forty-one detector assemblies are arranged as one row of five slats and nine rows of four slats inside the TOFp tray, which in total cover approximately one unit of pseudorapidity and $\sim 1/60^{th}$ of the full azimuth of the STAR TPC.

## *2.1 TOFp*

An exploded view of the TOFp detector is shown in Figure 3. The mechanical structure is a 50-mil thick welded aluminum box, called a "tray," of dimensions 8.5"×3.5"×95". Aluminum brackets are riveted to the bottom of the tray which slide onto rails attached to the outer field cage of the STAR TPC. This holds the detector on the exterior surface of the STAR tracking detector on the East side of STAR at approximately the 7 o'clock position. Inside the tray are the scintillator+PMT+base assemblies, the front-end electronics, the signal cabling, and aluminum tubing for heat removal via water flow.

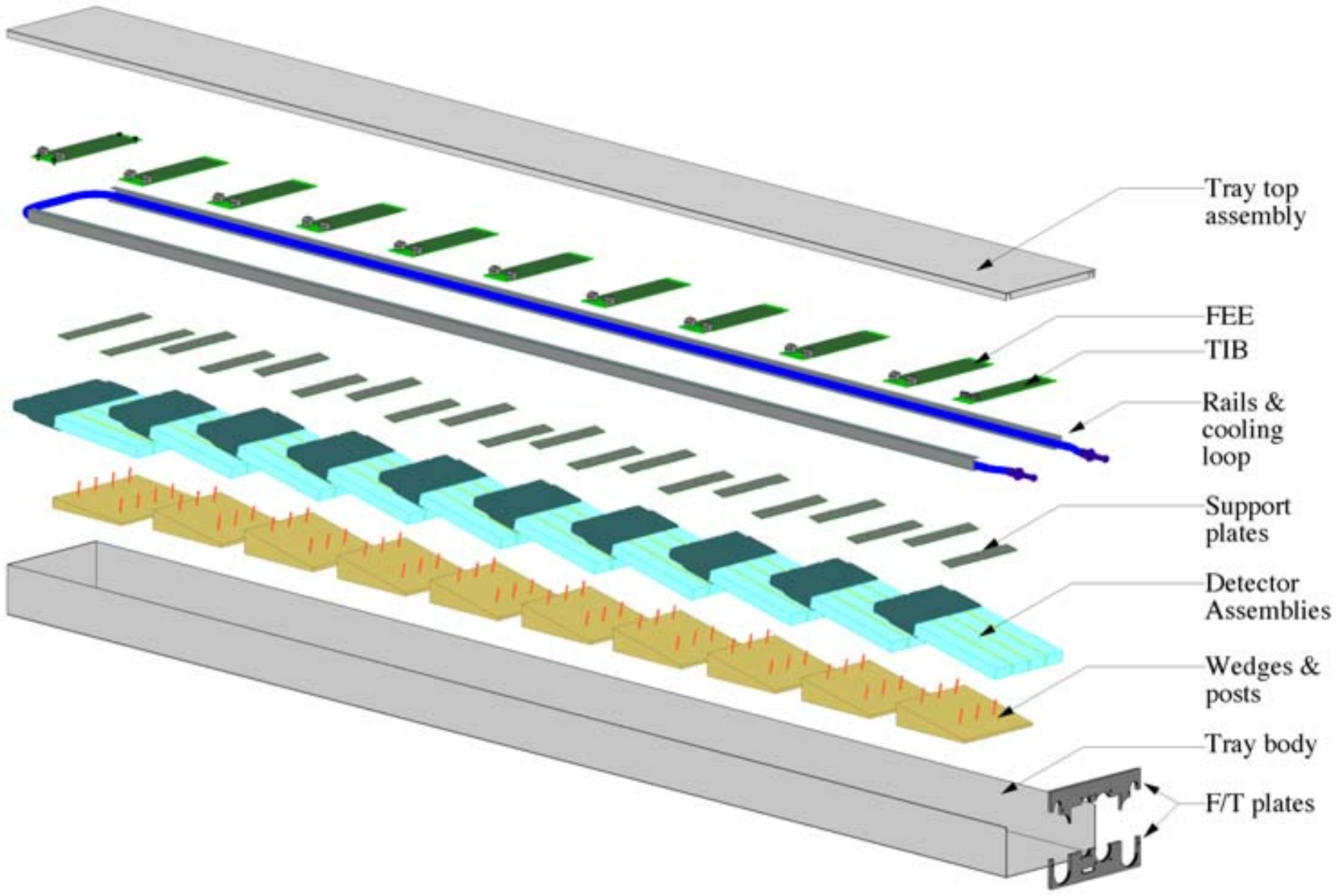

Fig. 3. The TOFp detector.

Each detector assembly consists of a 1.5"×2cm×20cm "slat" of Bicron BC420

plastic scintillator with diamond-milled edges. This formulation was chosen because of its fast fluorescence rise time of 500 ps. Epotek 301-2 spectrally-transparent epoxy is used to bond one end of a slat to a Hamamatsu R5946 Photomultiplier Tube (PMT). The slats are thus short and "single-ended." This epoxy was cured for ∼48 hrs at a temperature of ∼90-95 °F.

The Hamamatsu R5946 PMTs have sixteen stages of mesh dynodes which allow operation in the 0.5T STAR magnetic field [5]. Additional specifications on the maximum output rise time (2.5 ns) and the tube-by-tube gain variation (1:10) were required for this system. These PMTs are powered using a custom Cockroft-Walton system called "HVSys" developed by Astakhov *et al.* [6] This system consists of PMT bases, called "cells", one per TOFp PMT, and a rack-mounted System Module that provides the control and read-back functions and a 200V line to drive the Cockroft-Walton divider networks inside each cell.

The scintillator bar is wrapped in two layers of Tyvek 1055B UV-treated paper. The measurements by Stoll [7] indicate that Tyvek is more reflective than aluminum foil and aluminized Mylar for scintillation light (and also less abrasive), leading to larger numbers of photoelectrons for each hit and hence better timing performance. A layer of 2 mil-thick black photographer's plastic surrounds the Tyvek layers to optically seal the slat assembly. Black PVC shrink tubing is recovered over the slat-to-PMT glue joint.

Table 1
The pseudorapidity, $\eta$, ranges covered by the 10 rows of TOFp slats for collisions occurring at $Z_{vtx}$=0.

| Slat Row | No. of Slats | Angle (deg.) | $\eta$ range $Z_{vtx}$=0 |
|---|---|---|---|
| 1 | 5 | 5 | 0.0506 - 0.146 |
| 2 | 4 | 10 | 0.147 - 0.242 |
| 3 | 4 | 12 | 0.243 - 0.338 |
| 4 | 4 | 12 | 0.339 - 0.433 |
| 5 | 4 | 12 | 0.436 - 0.529 |
| 6 | 4 | 12 | 0.531 - 0.621 |
| 7 | 4 | 12 | 0.624 - 0.711 |
| 8 | 4 | 12 | 0.714 - 0.798 |
| 9 | 4 | 12 | 0.801 - 0.882 |
| 10 | 4 | 12 | 0.886 - 0.963 |

A total of forty-one detector assemblies are arranged as ten rows of slats inside the tray and, for collisions at Z=0 (the center of STAR), subtend the pseudorapidity ranges shown in Table 1. Each row of slats is held at a specific position and angle with respect to the tray bottom using wedge-shaped pieces of 3 lb/cu.ft. Last-A-Foam [8], a rigid closed-cell polyurethane foam. This foam produces dust when rubbed, so each wedge was coated with several layers of spray polyurethane before it was glued into the tray. Either eight (first row) or six (rows 2 through 10) Teflon 6-32 threaded rods are anchored to the underside of each wedge and extend upward to allow the detector assemblies to be bolted down onto the tops of the wedges. The five slats in the first row are at an angle with respect to the tray of 5 degrees, the second row slats are at 10 degrees, and the third through tenth row slats are at the maximum angle possible of ~12 degrees. This implies the first two rows of slats are generally projective to the collisions, while the remaining rows become less projective going across the tray out to $\eta \sim 1$. The average angle of incidence for tracks in the tenth row of slats is near 20-30 degrees, depending on the Z-location of the collision along the beam pipe.

The signal cables are routed along the bottom and sides of the tray and exit from the “feedthrough” end of the tray, which is shown near the bottom right of figure 3 with the label “F/T”. Above the signal cables and around three sides of the tray is a welded 1/2” O.D. aluminum tubing loop, which carries water from the STAR TPC cooling system at a rate of ~2 G/m. This tubing is mounted inside two 92”-long 1/16”-thick aluminum architectural channels which are machine-screwed to the two long side walls of the tray via aluminum nuts welded to the inside vertical sides of the channels. The front end electronics are mounted on the top side of the channels via aluminum machine screws that extend upward through the channels and are also welded in place. This provides a high-conductivity thermal path between the electronics boards and the cooling water.

The on-detector electronics for the TOFp tray are described in section 2.3.1 below. The resulting signals travel over the path described in sections 2.3.2 and 2.3.3, and are digitized in a CAMAC crate as described in section 2.4.

### *2.2 pVPD*

The pVPD consists of the two identical detector assemblies shown in Figure 4, one on each side of STAR very close to the beam pipe at a distance of |Z|=5.6 m from the center of STAR. The mounting structure consists of an aluminum base plate, two 1”-thick machined Delrin face plates, and three welded aluminum rails, to which detector elements are attached using pipe clamps. The base plate of the mounting structure clamps to the aluminum I-beam that is used to support the RHIC beam pipe. An aluminum box, labelled "FEE & TIB box" in this figure, attaches to the underside of the base plate for the on-detector electronics. It is electrically isolated from the rest of the pVPD structure since the I-beam and beam pipe are at an electrical ground other than that used by all STAR detectors and electronics.

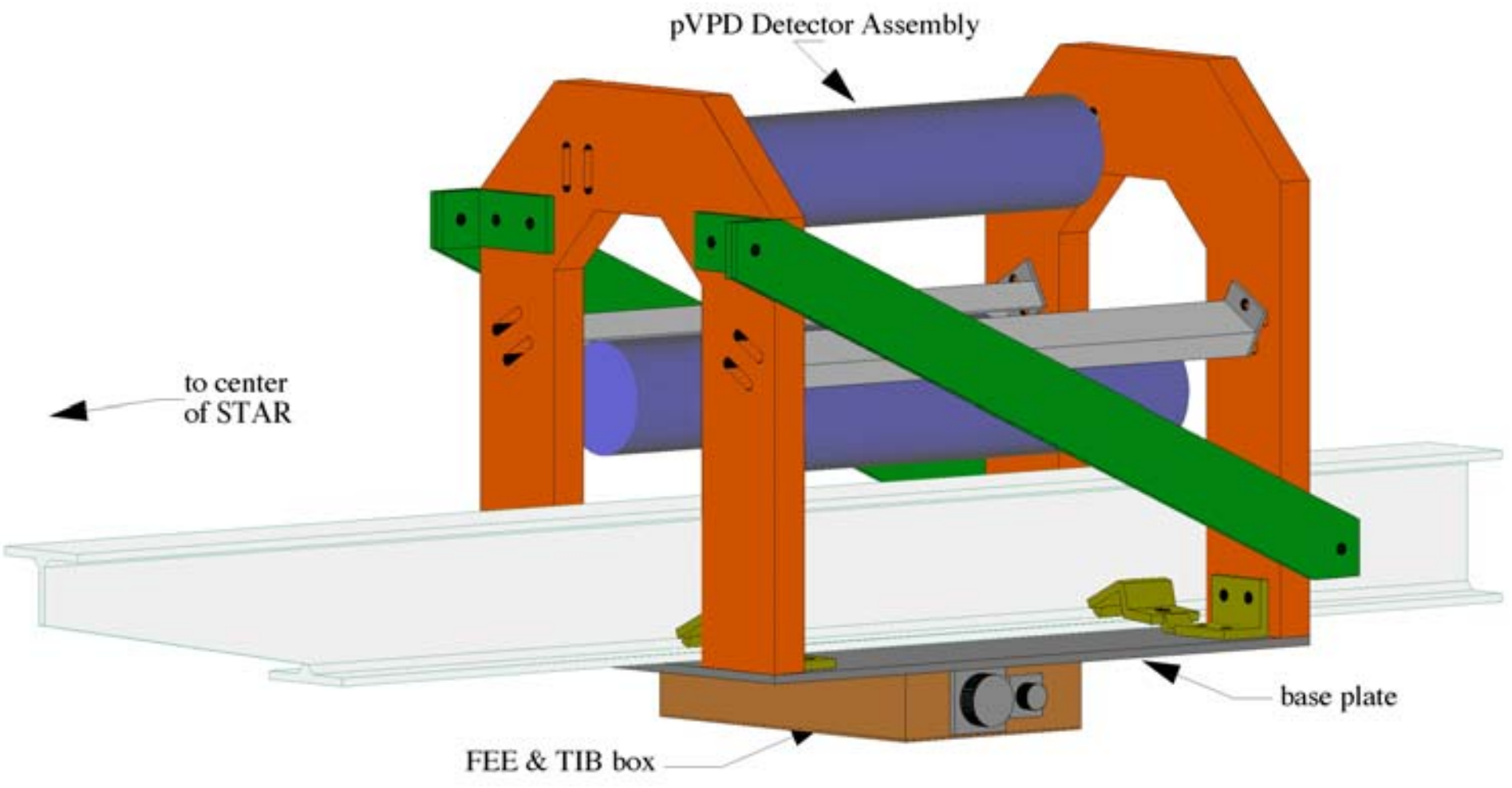

Fig. 4. One of the two identical pVPD detector assemblies.

The magnetic fringe field near the pVPD detectors when the STAR magnet is on is a few hundred Gauss, requiring that the pVPD PMTs be magnetically shielded, and that the mounting structure can withstand the magnetic forces. The three shielded detector elements in each detector assembly are held very close to the beam pipe at the 12, 4, and 8 o'clock positions. For collisions at Z=0, the active elements cover approximately 19% of the total solid angle in the pseudorapidity interval $4.43<\eta<4.94$, or laboratory polar angles in the range $0.82<\theta<1.48$ degrees.

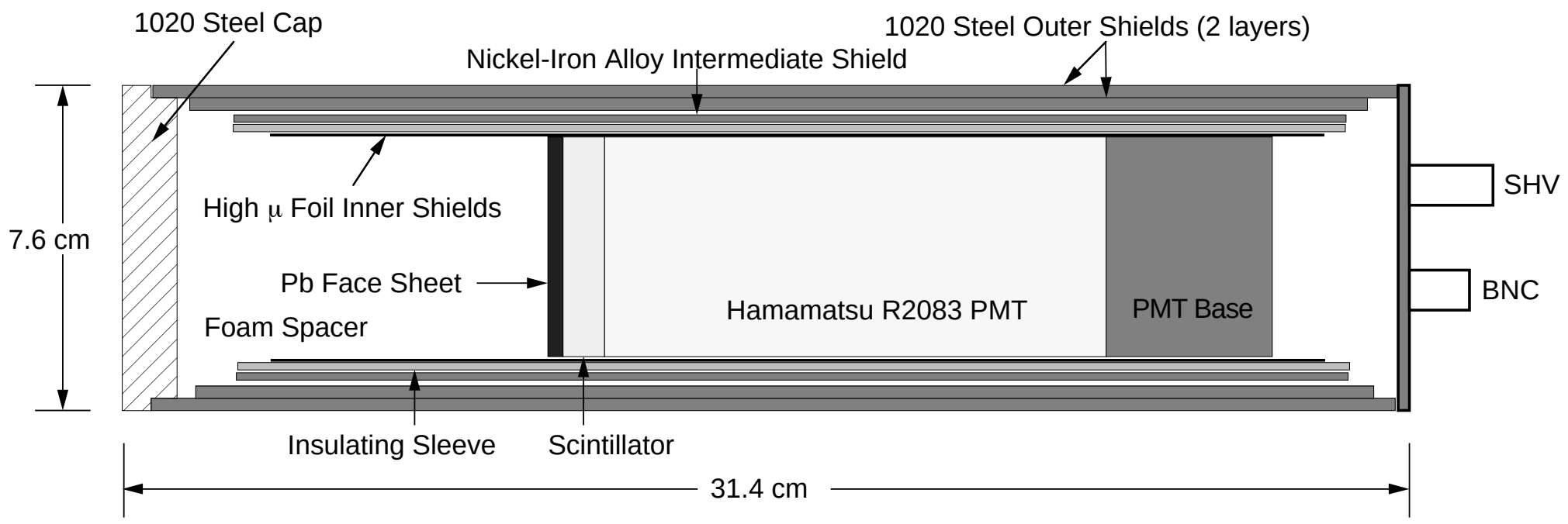

Fig. 5. A cut-away side view of a pVPD detector element.

Shown in Figure 5 is side view of one of the six pVPD detector elements. As seen by the incoming particles, each detector element consists of the front "cap" of the magnetic shield, an air gap, a ~1cm-thick layer of Lead (~1 radiation length), a 1/4"-thick layer of Bicron BC-420 scintillator, a Hamamatsu 2083 2" PMT, and finally a linear resistive base. These detectors are thus of the “flashlight” design. The lead acts as a photon converter, which floods the following scintillator layer with electrons and leads to large PMT signals and excellent start-timing performance. The high voltage for the pVPD bases is generated on the platform by a LeCroy 1440 HV mainframe and carried to the detectors over RG-59 coaxial cables. The weight of each detector element is approximately ∼10 lbs which comes almost entirely from the magnetic shielding.

The on-detector electronics for the pVPD are the same as those used in the TOFp tray (section 2.3.1). The pVPD signals travel over a similar signal path as for TOFp (see sections 2.3.2 and 2.3.3), and are digitized in the same CAMAC crate (see section 2.4). The starts, gates, fast clears, and test signals for this digitization are provided by a “local trigger” system constructed from NIM logic and which takes as detector input signals only those from the pVPD.

## *2.3 Signal Paths*

This section describes the signal processing, starting from the outputs of the PMT bases on the detectors to the inputs of the digitizers which are the ADCs and TDCs on the platform. This includes the Front-End Electronics (FEE) (section 2.3.1), the platform discriminators (section 2.3.2), and the signal cabling (section 2.3.3). An overview of the TOFp and pVPD signal paths is shown in Figure 6. On the left side of this figure are the detectors, and on the right side are the digitizers. The details on the components in between are presented in the following subsections.

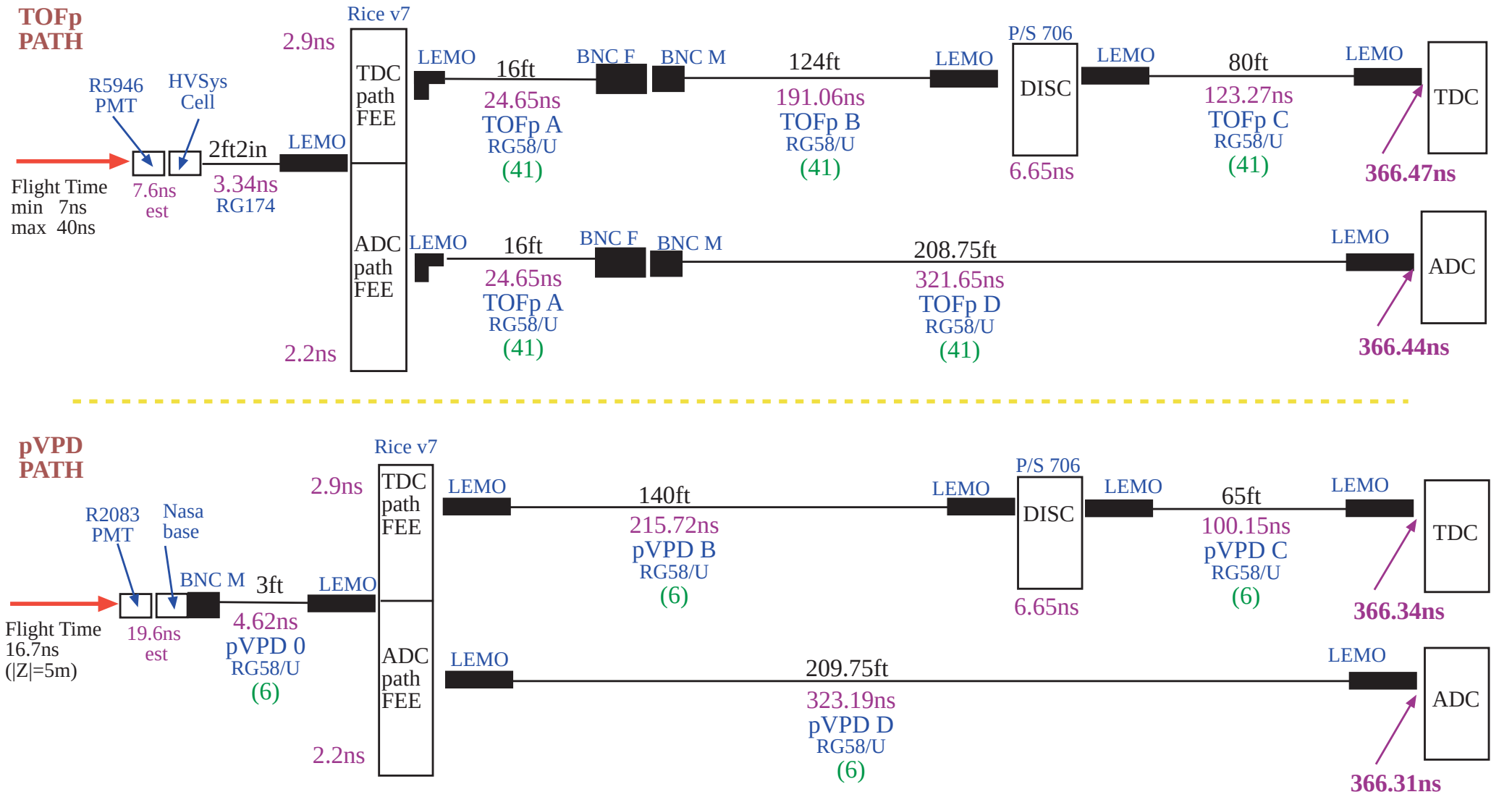

Fig. 6. The overview of the TOFp (upper) and pVPD (lower) signal paths from the detectors (left) to the inputs to the digitizers (right).

### *2.3.1 Front-End Electronics*

Custom Front-End Electronics (FEE) are mounted on the detectors. For each input channel, these produce a NIM-standard logic pulse from a fast discrimination of the detector signals as well as a copy of the raw input PMT signal. The logic pulse is used for the timing, while the copy of the raw PMT signal is used for the correction of the timing for the slewing effect [9] [10]. Leading-edge discrimination (LED) was implemented as it is electronically simpler than constant fraction (CFD) or crossover discrimination, and as it will also outperform them for the present detectors [11]. In general, when the scintillator is long or wide, the combined effects of reflections and absorptions inside the scintillator cause the pulse shape to vary with the distance between the particle hit and the PMT [12]. A correlation between the particles' average angle of incidence and their position along the length of the scintillator would also cause variations in the pulse shape with the distance to the PMT. CFDs outperform LEDs only when all of the pulses have the same shape [9] [11] [12]. An essentially complete correction for slewing is possible offline, *e.g.* as described in Ref. [10]. Care was taken during the development of these electronics to minimize the rise time of the timing logic pulse output.

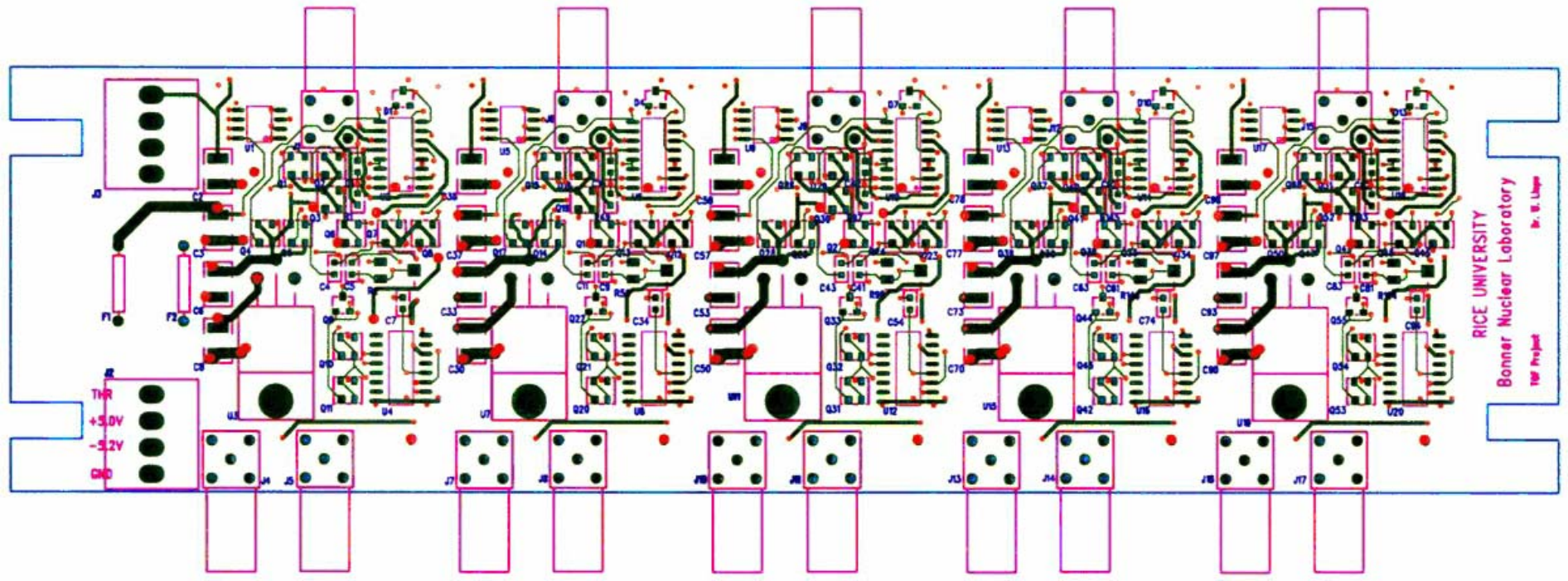

Fig. 7. The top view of the FEE used in both the TOFp and pVPD detectors.

A top view of a FEE board is shown in Figure 7. The size of the boards was constrained by the space available for mounting inside the TOFp tray. There is one FEE board per TOFp slat row, for a total of ten FEE boards inside the tray. The same boards are used without modification on the start side as well - one FEE board for each pVPD detector assembly, East and West. Each FEE board is a four-layer printed circuit board of dimensions 8.2"×2.5", and each can handle input signals from up to five detector channels. Seen on the left side of Figure 7 are the low voltage connectors and the fusing for over-current protection. The input signals enter the board on the top via five board-mounted Lemo jacks, while the outputs are also Lemo jacks and are seen on the bottom.

These boards perform a leading-edge discrimination relative to a remotely adjustable threshold voltage. The discrimination function is performed by an Analog Devices AD96685 Integrated Circuit voltage comparator. This comparator was chosen based on its specification of a low propagation delay dispersion (50ps) versus the input drive. The signal voltage developed in a PMT anode load resistor is applied directly to the comparator with only the addition of a protection diode network. This input protection clips input pulses larger than 4V, where the typical pulse height for the particle hits of interest is ∼0.5V. A fast NIM-standard logic signal of a constant width of 30 ns is generated by mostly discrete circuitry surrounding the comparator.

A view of the logic output pulse from these boards is shown in the upper left frame of Figure 11 in section 2.3.3 below. No attempt was made to smooth out the structure in the flat-top section of these output pulses, as this structure does not affect the signal timing which is based only on the pulses' leading edge. The main effort was rather the optimization of the rise time of these logic signals. The inherent 2 ns output rise time of the comparator output was reduced to ∼0.45 ns in the logic signal output of the FEE by careful use of frequency compensation and feedback. This should be compared to the output rise time for the fastest commercial NIM and CAMAC discriminators, which is ∼2 ns, a factor of four larger than that from the present boards. This fast output rise time leads to the best possible system performance in the presence of the long signal cables that are used to bring the TOFp and pVPD signals to the STAR platform for digitization. The timing resolution for PMT-like pulser signals is extremely good, being on the order of ∼15-20 ps as measured by an 8 GSa/s HP Infinium oscilloscope.

The input signal is also fed to an on-board I.C. buffer amplifier which isolates and drives the second output to be digitized in an ADC for the offline slewing corrections. The output gains of the buffer amplifiers were matched to 1% across all FEE channels.

The channel dead time per hit was in the range 60-70 ns, depending on the threshold. The absolute times of flight of particles striking the TOFp slats in STAR range from a minimum of 7 ns for v=c hits at $\eta\sim0$, to a maximum of 40 ns for the slowest measurable hadrons (500 MeV/c deuterons near $\eta\sim1$). One requires that each FEE channel to be ready for another input pulse before the next possible beam crossing. At RHIC, the time between crossings is 106 ns. Thus, the dead time of the FEE boards of approximately 65 ns is adequate for use in STAR.

For TOFp, the low voltage is provided to the on-detector electronics by Kepco 15-12M power supplies over 120 feet of 10AWG VN-TC cable. For the two pVPD detectors, the same type of cable is driven by a BK Precision Model 1780 dual-output power supply. The required low voltage inputs are +5V and -5.2V, and the current drawn by each FEE channel is approximately 250 mA and 850 mA, respectively.

The externally-supplied, positive polarity, external threshold voltage input is inverted and distributed to all channels of the FEE. The Remote Threshold System (RTS) provides the means to accurately control and read-back the FEE threshold level over long cables from the platform. An analog current signal transmission (4-20 mA loop) technique was chosen because of its simple implementation and proven reliability in a wide range of industrial and process control environments. The control signal current source is galvanically isolated from the discriminator power system to further eliminate the possibility of ground loop induced error. The operator interface at the platform consists of a 4U rack mounted control panel with precision 10 turn dial potentiometers and four digit displays for each of three remote thresholds (TOFp, pVPD East, and pVPD West). The rack unit connects to a Threshold Interface Board (TIB), one mounted on each detector near the FEE boards. The low voltage needed by the TIBs is derived locally from that delivered to the FEE.

The RTS includes a redundant measurement of the control signal at the detectors, a fail-safe default to a fixed preset threshold in the case of any failure, and a "read-back" indication of the actual applied threshold voltage. The fail-safe function is enabled under either of two fault conditions. First, when a possible failure of the current receiver circuit on-detector is suggested by a difference between its output and that of a redundant receiver. Second, if the control current falls significantly below the zero scale value of 4 mA, which might be due to a failure of the connecting cable or the control panel power supply. The facility to pre-define the default FEE threshold, typically 50 mV, is provided by a trimmer type potentiometer on the each TIB on-detector. None of these failure modes have been observed.

### *2.3.2 Platform Discriminators*

To get the signals off the detectors, and to allow time for a local trigger decision (section 2.4), coaxial cables are used to propagate and delay the FEE output signals (section 2.3.3) as shown in Figure 6. To obtain the best possible rise time of the timing signals at the input to the digitizers, the common technique [13] [14] of "line driving" was used. The timing signal path after the FEE thus consists of a section of cable to bring the signals off the detectors ("B-section" in Figure 6), a re-discrimination of the timing signals on the platform, and finally another section of cable ("C-section" in Figure 6) to allow time for the formation of the local trigger decision and to input the timing signals to the digitizers. The rise time of the logic signals at the input to the digitizers is approximately a factor of two smaller than if the intermediate discrimination was not employed, as discussed in section 2.3.3 below.

Given the 20-35% detector occupancies in full-energy central Au+Au collisions, the timing cross-talk performance of the commercial discriminators used on the platform as line drivers is thus highly relevant. It is very difficult to correct for timing cross-talk in the off-line data analysis, so this effect must be suppressed by design.

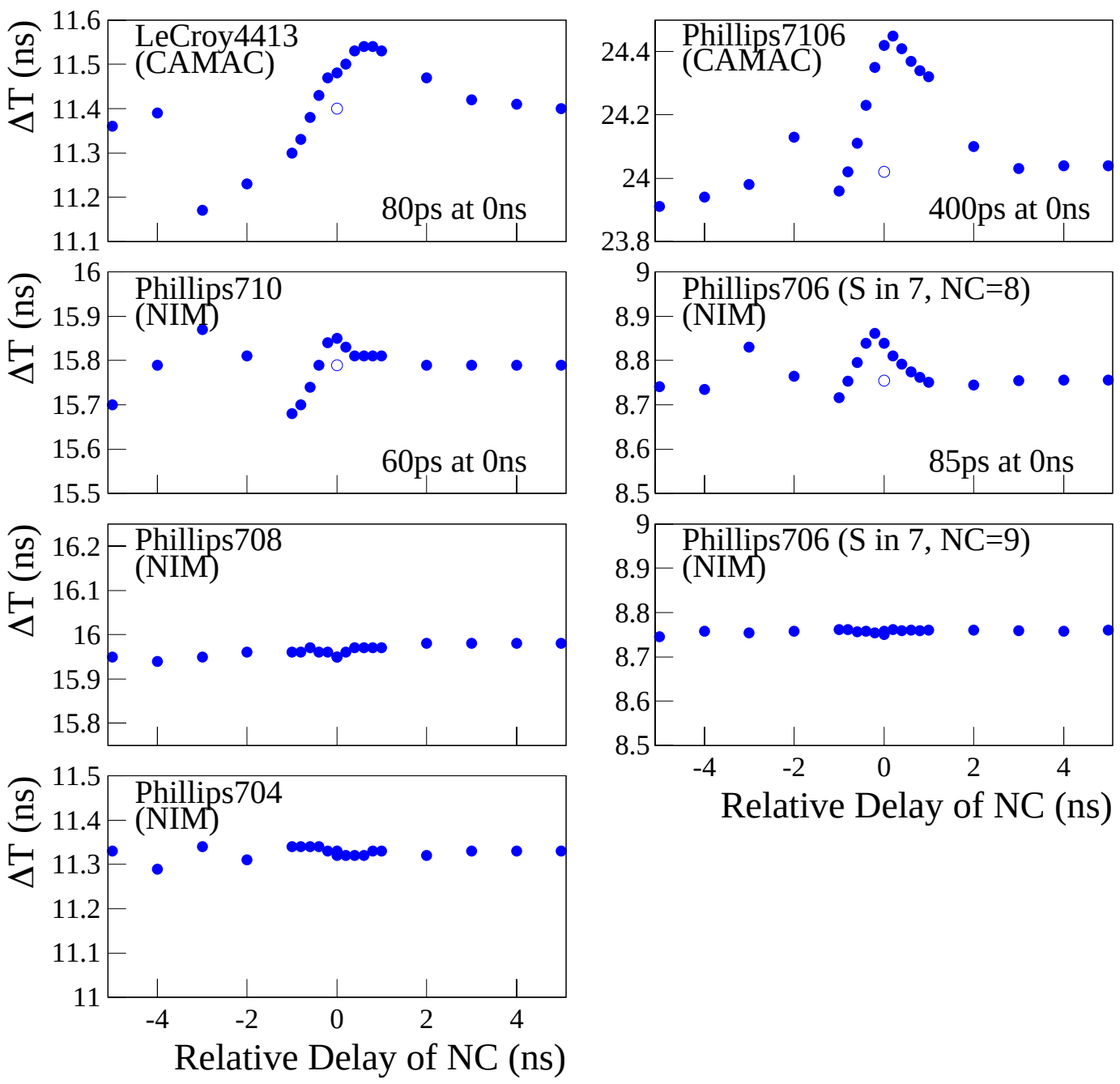

Fig. 8. The measured timing cross-talk for a number of common CAMAC and NIM discriminators. For the details, see the text.

A number of commercial NIM & CAMAC discriminators suitable for use as the TOFp and pVPD line drivers were tested for timing cross-talk. Shown in Figure 8 are the results, where each frame corresponds to a specific manufacturer and model. The upper two (LeCroy 4413 and Phillips Scientific 7106) are 16 channel CAMAC units, while the other five are Phillips Scientific NIM units - the PS704 (4 ch), PS710 (4 ch), PS708 (8 ch), and PS706 (16 ch). The vertical axes depict the time interval measured in an 8 GSa/s HP Infinium oscilloscope and defined by pulsing one channel of one of these discriminators, the "channel under study" (CUS). The open points in Figure 8 indicate the (arbitrary) value of the time interval when there are no signals in any of the channels that are adjacent to the CUS. The solid points depict the values of the same time intervals when there is a signal input to a "neighboring channel" (NC), while the horizontal axes are the relative delay between these two signals. In the absence of timing cross talk, a channel's response is unaffected by near-time pulses in neighboring channels. This would mean the solid point at a relative delay of zero nanoseconds should lie on top of the open point (no signal in any NC), and that the trend of the solid points for positive or negative relative delays should be flat.

This is clearly not the case for either of the CAMAC discriminators, where the modification to the pure start/stop timing of a single channel can be modified by *hundreds* of picoseconds if there are near-time signals in neighboring channels. The effect extends for a long relative time on either side and, even worse (but not shown in this figure), similar effects are not limited to only the nearest neighbor pairs of channels. For nearest neighbor pairs, the PS710 and PS706 also show timing cross-talk but it is relatively smaller and does not extend as far on either side in relative delay. The PS704 and PS708 show no measurable timing cross-talk for any pairs of channels (at the full 8 GSa/s bandwidth, the resolution of the Infinium is $\sim$15 ps). The lowest frame on the right side of Figure 8 shows the results for the PS706 unit when only every other channel of the unit is allowed. The only options free from timing cross-talk were thus the PS704 and PS708 units, or the PS706 unit when using only every other input. We chose the last option since these units were readily available from the BNL HEEP equipment pool.

### *2.3.3 Signal Cabling*

Approximately 97% of the $\sim$340 ns of total delay along the signal paths between the detectors and the digitization is provided by coaxial signal cables. We compared different varieties of coaxial cable which are all 50$\Omega$ impedance but differ in the manufacturer's specification of the cables' attenuation (dB) versus frequency (MHz). Coaxial cable with a relatively low attenuation at high frequencies (>200 MHz) will degrade the rise-time of the timing signals the least, leading to the best possible timing performance at the digitizers. Based simply on various manufacturers' specifications for the cable attenuation per unit length at high frequency, we chose Belden 9310 as the signal cable used throughout both the TOFp and pVPD signal paths. A comparison to other similar cables, and parameter-free simulations of the performance of the Belden 9310 cable for the present system, are described in this section.

We measured the 0-100%-equivalent rise time in nanoseconds for the logic signals from a PS706 discriminator (output rise time is 2.1 ns) plus a specific length of cable. The rise-times were measured in the 10%-30% region of the output pulse heights, *i.e.* the fastest part of the leading edges where the thresholds will be placed for the best timing performance. These values are then multiplied by a factor of five to give the 0-100%-equivalent rise times that are shown in the figure. Shown in Figure 9 is this output rise time versus the cable length in nanoseconds for different varieties of 50$\Omega$ coaxial cable, as labelled. Belden 8259 was evaluated also in Ref. [15], and Belden 8219 is used in the PHENIX TOF system [16]. According to the figure, the degradation of the logic signal rise time with the length of the signal cable is approximately linear for all cable types. The rise time degradation for the Belden 9310 cable is significantly less than that from all the other cables studied, over the entire range of cable delays. Over 200 ns of delay, the Belden 9310 rise times degrade from $\sim$2.1 ns to $\sim$2.7 ns, while the other cable types degrade the rise times to ∼3.7 ns or higher. The improvement of the rise times of the logic signals after the cable delay by choosing Belden 9310 naturally improves the timing resolution of the digitizers [14].

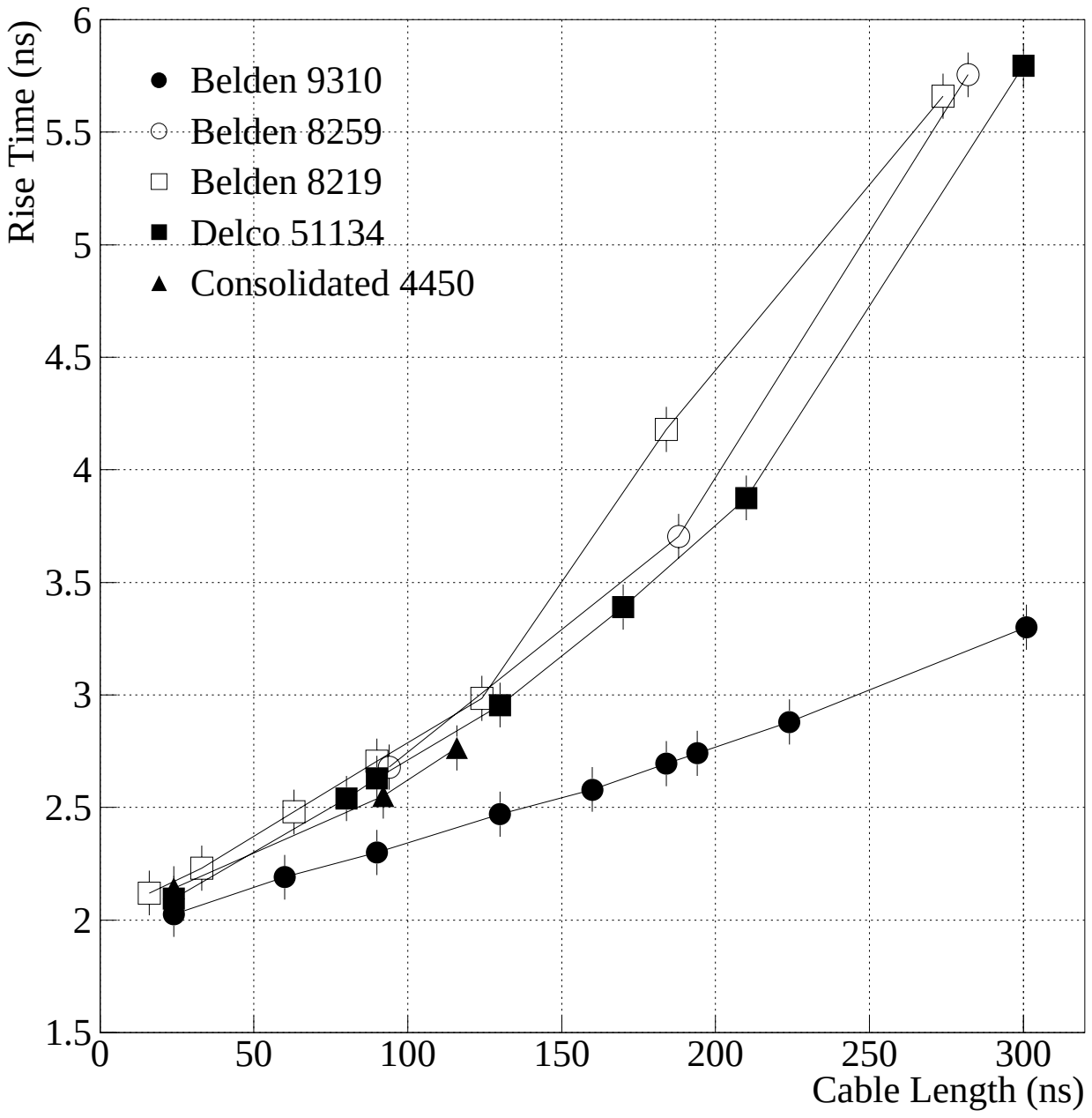

Fig. 9. The 100%-equivalent rise time (in ns) of the logic signals from a PS706 discriminator versus the length of cable (in ns) for different varieties of RG-58 coaxial cable.

To further explore the performance of the Belden 9310 for the present system, we compared measurements of the analog and timing signals at various points along the TOFp signal paths to those predicted by a model [17,18]. An arbitrary input pulse with voltage versus time, $V(0,t)$, is modified by the transmission over a cable of length, $L$, into the output pulse $V'(L,t)$. The model involves the Fourier transform of the input pulse,

$$F^*_{input}(\omega) = \int_{-\infty}^{\infty} V_{input}(0,t)\ e^{-i\omega t}\ dt, \tag{1}$$

the application of the (complex) attenuation versus frequency curve, $\alpha^*$, for a given cable type,

$$F^*_{output}(L,\omega) = F^*_{input}(\omega)\ e^{-\alpha^* L}, \tag{2}$$

and finally the inverse Fourier transform provides the model's prediction of the output pulse,

$$V'(L,t) = \frac{1}{2\pi}\int_{-\infty}^{\infty} F^*_{output}(L,\omega)\ e^{-i\omega t}\ d\omega. \tag{3}$$

The transforms are performed with the routine RFFT from the CERN program library [19]. The dimensionality of the space used by this routine was increased from the $2^8$ used in Ref. [18] to $2^{14}$ to improve the simulation of fast rise time signals. The cable transmission increases pulse widths and rise times, and decreases pulse heights, to extents depending only on the length of the cable and the cable's attenuation curves. The complex attenuation function is parameterized using

$$\alpha^*(f, Z) = [c_0 + c_1\sqrt{2if} + c_2 f]_Z \tag{4}$$

while the real part of $\alpha^*(f, Z)$ and the manufacturer-specified attenuation, $\lambda$, in dB, are related by,

$$\mathrm{Re}(\alpha^*) = \frac{\log(10)}{20}\,\lambda[\mathrm{dB}]. \tag{5}$$

The constants $c_0$, $c_1$, and $c_2$ are thus fit parameters [20] for a given cable type and unit length which completely define how this cable type will modify arbitrary input pulses. The constants are shown in Table 2 for the Belden 9310 and Belden 8259 cables. A good agreement between the present values and those from Ref. [18] for the Belden 8259 is observed.

Table 2
The coefficients for the complex cable attenuation function, $\alpha^*$, for the Belden 8259 and Belden 9310 cable types.

| | $c_0$ | $c_1$ | $c_2$ | |
|---|---|---|---|---|
| Belden 8259 | 0.035 | 0.144 | 0.0036 | from Ref. [18] |
| Belden 8259 | 0.0348 | 0.143 | 0.00362 | this work |
| Belden 9310 | 0.115 | 0.117 | 0.00140 | this work |

Using the present parameters for the Belden 9310 cable, the model predictions for the pulse shapes at various points along the TOFp cable paths were compared to direct measurements as shown in Figures 10 (analog path) and 11 (timing path). The points are the measured wave forms obtained from the Infinium oscilloscope (only every $4^{th}$ point is drawn in the figure for clarity). The solid lines are the model predictions when using the waveforms in the leftmost frames as the input pulse shape, $V(0,t)$ and for the cable lengths as labelled (see also figure 6). The model accurately reproduces the pulse heights, widths, and rise and fall times along both the analog and timing signal paths without any free parameters.

Shown in the lower middle frame of Figure 11 is the model prediction if the PS706 "line drivers" (see section 2.3.2) were to be removed from the signal paths. In this case the timing signal rise times at the input to the digitizers degrades from $\sim$2.3 ns to $\sim$4 ns, underscoring the need for the line drivers for this system.

The various cable types seen in Figure 6 were fabricated using an odometer-equipped cable cutting machine which was able to control the various cable lengths to approximately 1 part in $10^3$. This implies hundreds of picoseconds of variation in the cable delays across different detector channels in the system which would contribute to the scatter of the offline timing offsets. However, this scatter was reduced drastically by matching, *e.g.* "B-section" cables that were slightly long(short) with "C-section" cables that were slightly short(long). After these assignments the standard deviation of the total delay in the B+C cables across the different channels in the TOFp and pVPD timing paths was reduced to $\lesssim$25 ps.

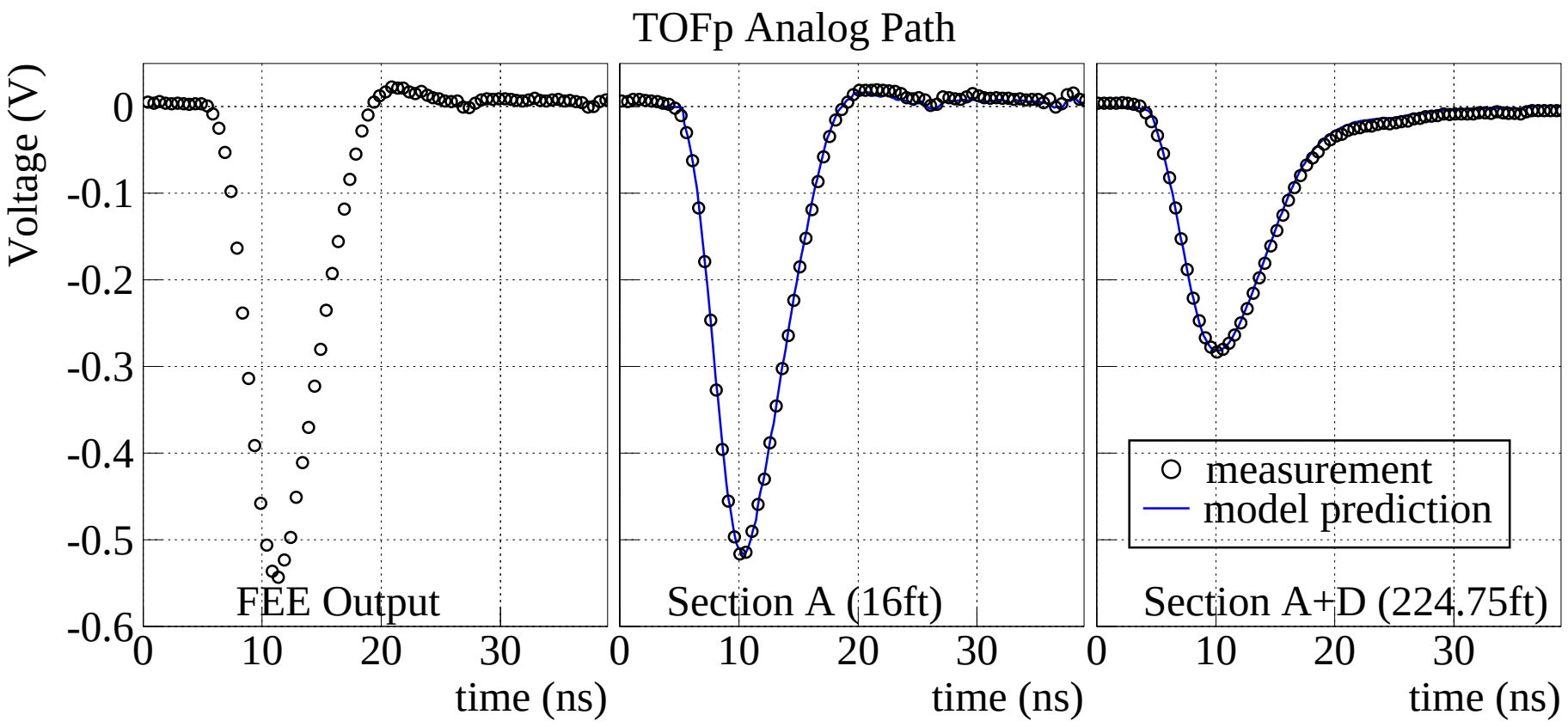

Fig. 10. The comparison of analog path signals measured by an oscilloscope (black points) and predicted by the model (solid lines) at the three stages of the signal propagation - FEE output (left frame), after A-section (middle frame), and at the ADC input after the A+D sections (right frame).

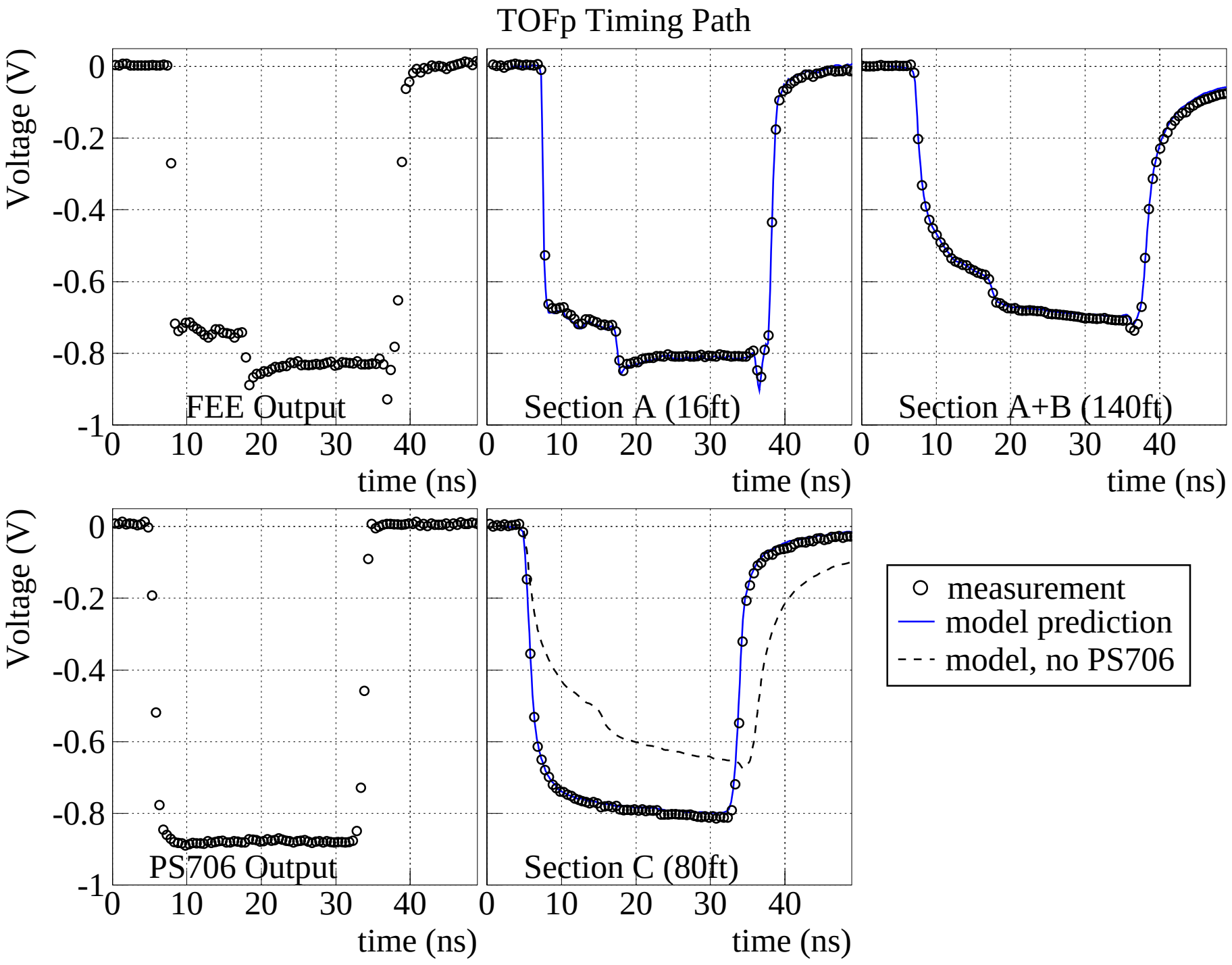

Fig. 11. The comparison of timing path signals measured by an oscilloscope (black points) and predicted by the model (solid lines) at the five stages of the signal propagation - FEE output (upper left frame), after A-section (upper middle frame), after the A+B sections (upper right frame), after the platform rediscrimination (lower left frame) and at the TDC input (lower middle frame).

### *2.4 Triggering and Digitization*

#### *2.4.1 Local Trigger System*

The STAR Trigger (TRG) system [21] identifies which beam crossings contain a collision event of interest, and which STAR detectors are live and are to be read out for this collision. This information is communicated to the STAR Data Acquisition (DAQ) system [22], which then collects the data from the list of detectors specified by TRG, and compiles and transmits these data to the RHIC Computing facility for archival and off-line analysis. During the 2001 RHIC Au+Au run, the information upon which the TRG system based its event selection was that from the detectors called the Central Trigger Barrel (CTB) and the Zero Degree Calorimeters (ZDCs). The CTB measures the charged particle multiplicity over the cylindrical radius of the TPC, covering two units of pseudorapidity, $|\eta|<1$, and $2\pi$ in azimuth. One of these CTB trays was removed and replaced with the present TOFp tray. The ZDCs measure the neutron multiplicity in a small solid angle near zero degrees with respect to the RHIC beam axis. The decision by TRG to initially accept a particular collision event is called a STAR "Level-0" trigger. When a collision is accepted at Level-0, each STAR detector designated to participate in the event is notified and told to identify this event with a 12-bit word called the "token". This notification occurs no later than 1.5 $\mu$s after the collision. This is an order of magnitude too long to consider using STAR Level-0 triggers to initiate the digitization of the TOFp and pVPD data. The signals from the TOFp and pVPD detectors are available on the platform ~240 ns after the collision, and it is not practical to delay the digitization of these signals for the additional ~1.3 $\mu$s needed to check if this event is later accepted by the STAR Level-0 trigger. Thus, the digitization of the TOFp and pVPD data is begun earlier, after a local decision based on the information from the pVPD itself. As the pVPD, CTB, and ZDC detectors cover different non-overlapping pseudorapidity regions, and as there may be gates placed on the CTB and ZDC information for Level-0 decisions by the TRG system, it is possible that collision events that fire the pVPD may not be later accepted by the Level-0 trigger, and *vice versa*. In order to properly interface with the STAR TRG and DAQ systems, despite these possibilities, a "Local Trigger" system for the TOFp and pVPD digitization was implemented in NIM logic and is described in this section. Also described in this section is the CAMAC hardware and Linux software used to locally read out the resulting TOFp and pVPD data and properly transmit it to the STAR DAQ system.

The schematic of the TOFp/pVPD Local Trigger system is shown in figure 12. The detector inputs, each of the six detector elements of the pVPD, are shown on the left side of this figure. On the right side of the figure are the various interfaces between the local trigger and acquisition systems and the STAR TRG and DAQ systems. The "line driver" discriminators (see section 2.3.2) have two outputs for each detector channel. For the six channels of discriminator on the pVPD timing path, one output drives the "C-section" cable (see figure 6) which brings the pVPD timing signals to the digitizers, while the other output is sent to the local trigger system as an input. The three pVPD detector elements on each side of STAR are sent to two layers of coincidence units. In the 2001 RHIC Au+Au running, these were set so that the local trigger required that at least two pVPD detectors on the East fired, and at least two pVPD detectors on the West fired, in order to begin the digitization.

The timing information from the TOFp and pVPD detectors is digitized in CAMAC by LeCroy 2228A Time to Digital Converters (TDCs), while the pulse area information is digitized in LeCroy 2249A Analog to Digital Converters (ADCs). The former is used in offline analyses to define the time intervals of interest, while the latter is used to correct the timing information for the slewing effect. The TDC's time conversion is 50 ps per TDC bin over a 11 bit dynamic range (100 ns full scale). The ADC conversion is 0.25 pC per ADC bin over a 10 bit dynamic range.

If the local trigger fires, gates and starts are delivered to each ADC and TDC, respectively, and the digitization begins. If approximately 1.3 $\mu$s later it is seen that the STAR Level-0 trigger also fired for this event, the token information is accepted from the STAR TRG system and locally associated in software with the event data presently being locally digitized. Approximately 100 $\mu$s later this digitization is complete. The data is then locally read out of the ADCs and TDCs by a Linux computer running custom software, reformatted to include the assigned token number, and then transmitted to the STAR DAQ system. If, however, the STAR Level-0 trigger did not accept the event that is in the process of being locally digitized, fast-clear signals are sent to each ADC and TDC, which terminates the digitization. If this occurs, the TOFp system is live and ready for another event approximately 2 $\mu$s later. For typical STAR data rates of $<$100 Hz, such fast clears have a neglibible effect on the live time of the system.

In the case that STAR has accepted an event at Level-0 but the pVPD, and hence the local trigger, did not fire, the present system must respond to STAR DAQ's request for TOFp and pVPD event data even though none exists for this event. In this situation, the local trigger system sends gates to the ADCs and test inputs to the TDCs, and the local digitization begins. The resulting data that is sent to STAR DAQ for these "events" is thus calibration data on the ADC pedestals and timing offsets internal to the TDCs. This intersperses calibration data with normal experimental events, which is useful to track slow variations in the system. Dedicated calibration runs are also taken several times per day as part of normal STAR shift crew operations.

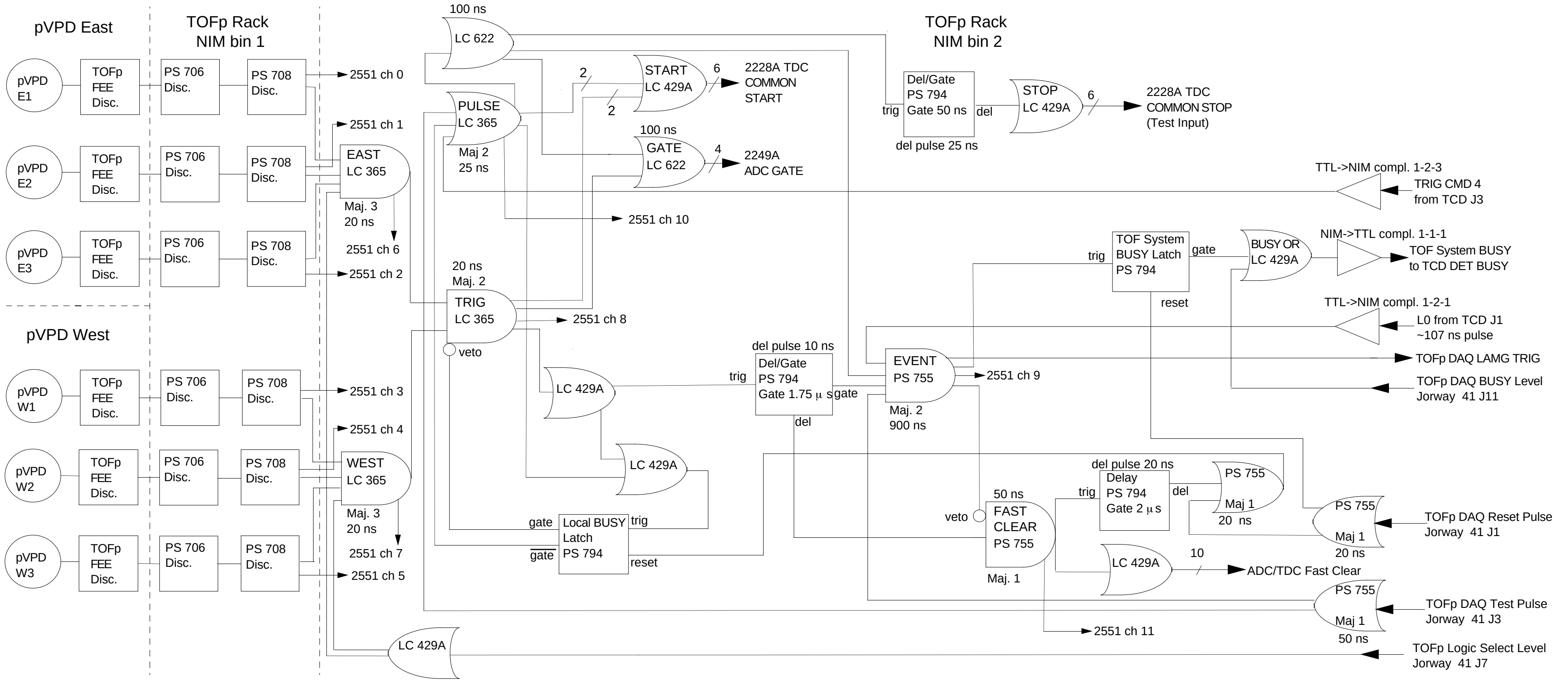

Fig. 12. The schematic of the TOFp/pVPD Local Trigger system.

### *2.4.2 Local DAQ System*

The local DAQ system controls the digitization hardware and reads out the detector data, reformats the data and sends it properly to the STAR DAQ system, controls I/O modules to properly communicate with the TOFp Local Trigger and the STAR Trigger, and visualizes the raw data. It is a conventional approach based on front-end electronics that are NIM and CAMAC modules with control by a PC running the Linux operating system. An overview of the local DAQ system is shown in Figure 13.

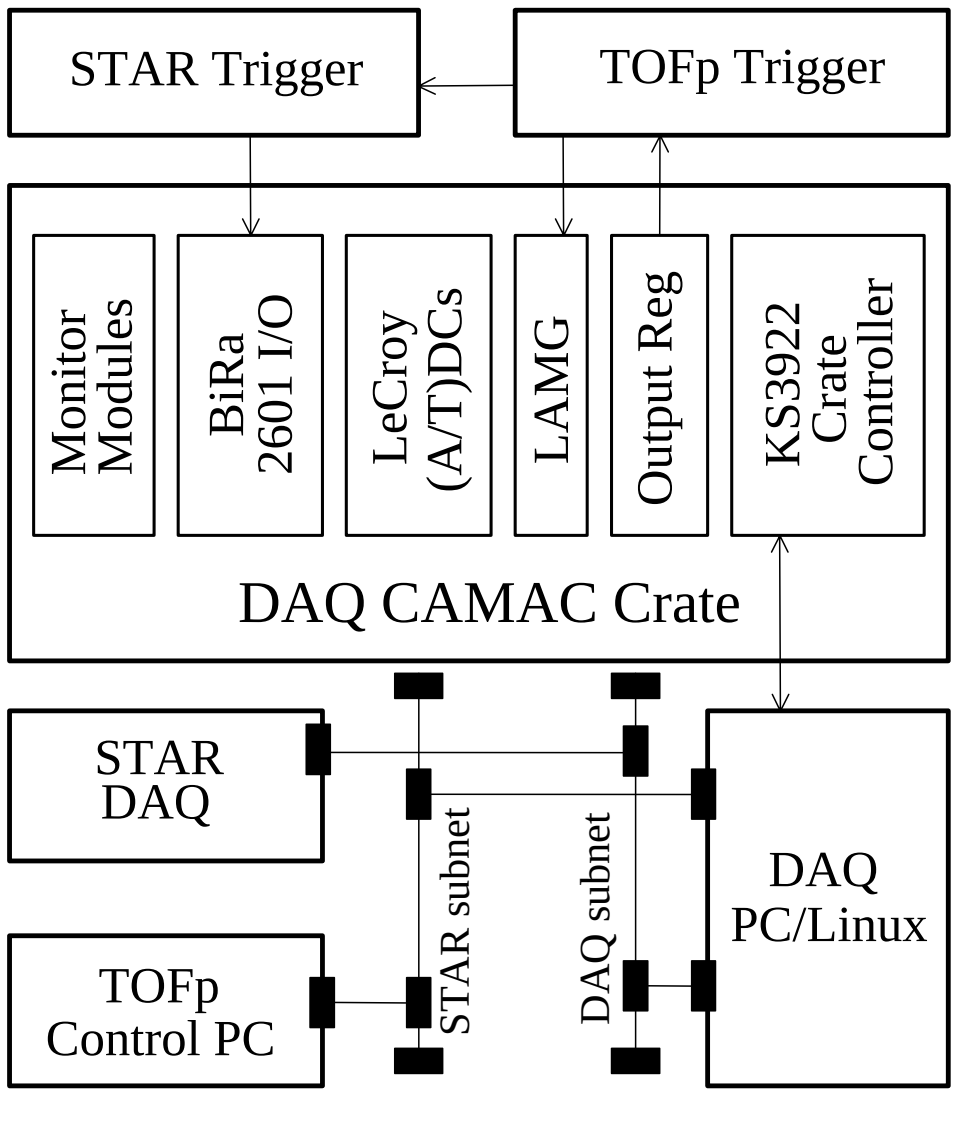

Fig. 13. The schematic of the hardware components of the local DAQ system.

There are four LeCroy 2249A ADCs and six LeCroy 2228A TDCs in the CAMAC crate. A Kinetics Systems 3922 parallel bus crate controller communicates with a Kinetics 2915 PCI Interface in the PC's PCI bus. The local DAQ system also needs to communicate with both the local trigger system and the STAR trigger. A Bira model 2601 24-bit I/O register accepts the token and other control information from the STAR Trigger, while a Jorway 41 output register and a custom "LAMG" module handle the communication with the local trigger system. The LAMG module is a single width CAMAC module which can accept a trigger signal and generate a LAM signal on the CAMAC bus after a preset delay. The detector and cable path temperatures, as well as the read-back values of the remote threshold system, are read out using a Kinetics model 3516 32-channel scanning A/D converter. A spare ADC channel is also used as a 60Hz clock as described in Ref. [23].

The data readout and transfer is done by the PC with a Kinetics model 2915 PCI card and two network interface cards (NIC). The function of the KS2915 is to communicate with the CAMAC crate and read out all the data on the CAMAC backplane. One of the NICs is used to communicate with the STAR DAQ system on DAQ's private subnet, while the other is used for monitoring and operator control via the general STAR subnet.

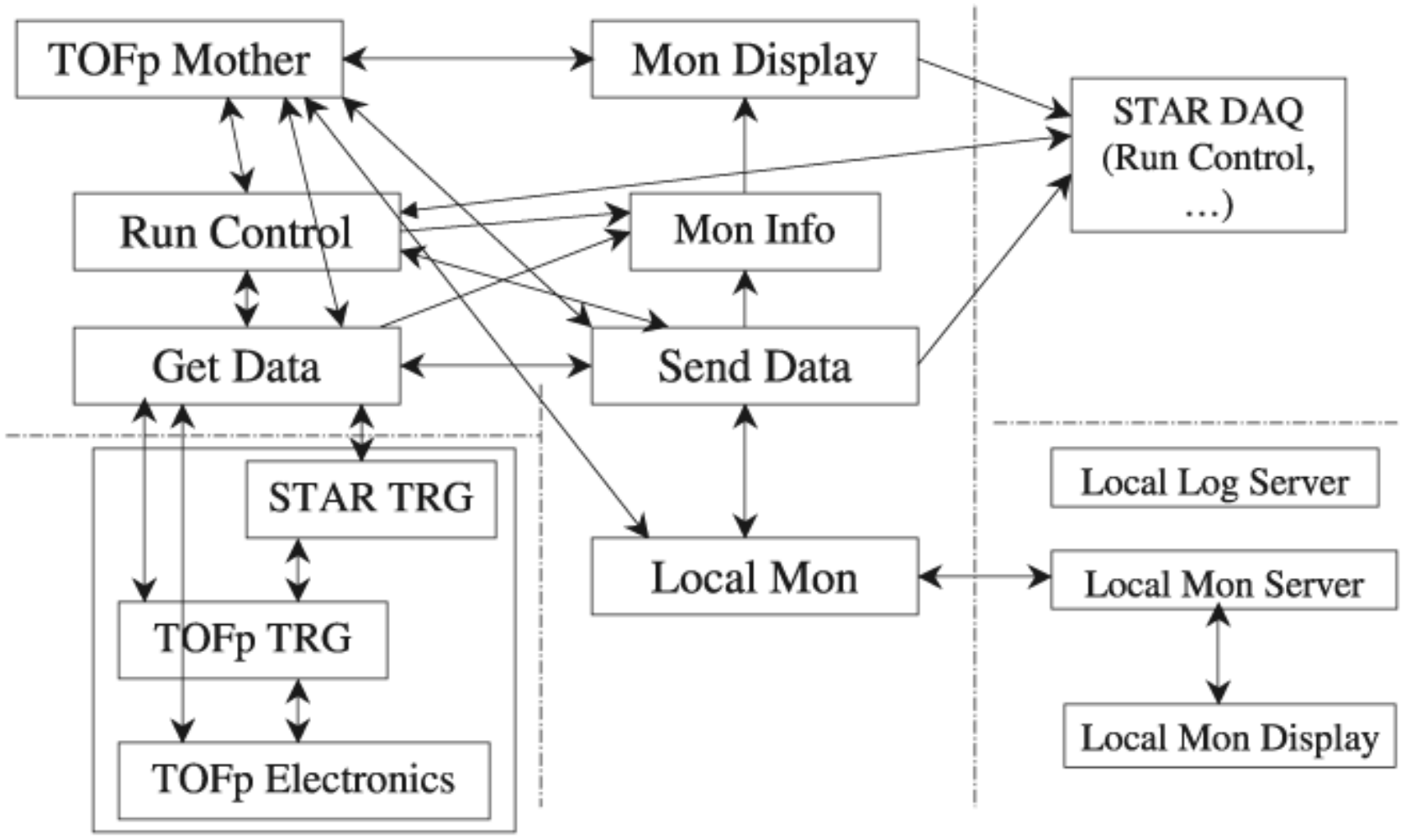

Fig. 14. The schematic of the software components of the local DAQ system.

Custom software written in C performs the functions outlined in Figure 14. There are five programs running on the PC. To simplify the operate procedure, they are spawned by one program called TOFPMOTHER. The five programs are RUNCONTROL, GETDATA, SENDDATA, STARDAQMONITOR and LOCALDAQMONITOR. RUNCONTROL communicates with STAR DAQ, gets the DAQ command, and notifies the other programs, then returns the execute status to STAR DAQ. It also updates generic status information for the system. GETDATA reads out the data and saves it in a shared buffer, so SENDDATA can send it to the STAR DAQ. GETDATA communicates with both the STAR Trigger and the TOFp local trigger system. In case the event buffer becomes full, GETDATA will keep and hold the system busy, so the STAR Trigger won't send further trigger information to TOFp. Each time an event arrives or is sent out, GETDATA and SENDDATA update the information stored in the shared memory segments. A round-robin scheduling algorithm is applied, so these programs have a higher priority than all other programs. STARDAQMONITOR sends run-time status information of the TOFp DAQ system to the STAR DAQ run-time system. LOCALDAQMONITOR sends a fraction of the raw data, typically every tenth event, to the local monitoring system. This allows the display of histograms of the raw data as it is arriving for diagnostics. The LocalDAQmonitor system is running on another PC in the STAR control room called the ControlPC. A local monitor server program runs as a daemon and receives the data from the local DAQ system, and saves it into histograms in a shared mapped ROOT [24] file. Another program called local monitor display displays these histograms.

## 3 Performance

### *3.1 pVPD*

The start detector for the TOFp System is the pVPD. The information needed from this detector is the corrected average of the times measured in those channels that fired in a given event. In the central Au+Au data from the 2001 run, all six channels of the pVPD fire in every event. The average of the six raw TDC values in each event is the crudest measure of the event start time, as it is smeared by the slewing effect intrinsic to PMT-based detectors. The first calibration necessary is thus the correction of the pVPD start times for the slewing effect.

In order to avoid bias during this correction, we compare the TDC average of one East pVPD channel and one West pVPD channel to the average of the other four pVPD TDC channels. This difference of averages, referred to as "$\langle 2\rangle - \langle 4\rangle$", has a standard deviation that is increased by the fact that all six of the times used in this average depend on the respective six ADC values. An iterative technique was employed where two channels are selected, *e.g.* East 1 and West 1, and the difference of TDC averages $\langle 2\rangle - \langle 4\rangle$ is plotted versus the two ADC values for East 1 and West 1, respectively. This two-dimensional profile is then fitted with a polynomial function of the two ADC values, and these fit parameters are used in later passes. The next step of the iteration involves the next pair of channels, *e.g.* East 2 and West 2. The difference of TDC averages, $\langle 2\rangle - \langle 4\rangle$, is plotted versus the ADC values for East 2 and West 2, respectively, where the $\langle 4\rangle$ average includes the corrected TDC values for East 1 and West 1 following the first iteration and the still uncorrected East 3 and West 3 TDC averages. This process is continued in for up to nine passes such that each East and West pair of pVPD TDCs is fit and corrected versus their respective pairs of ADC values up to three times.

The results from this process are shown in figure 15. The different point styles in this figure correspond to the first four passes as labelled, after which the correction is essentially complete. The highest points are the raw data, and have a resolution on the quantity $\langle 2\rangle - \langle 4\rangle$ of ~2.6 TDC channels, or ~130 ps. After the iterative procedure, the standard deviation of the quantity $\langle 2\rangle - \langle 4\rangle$

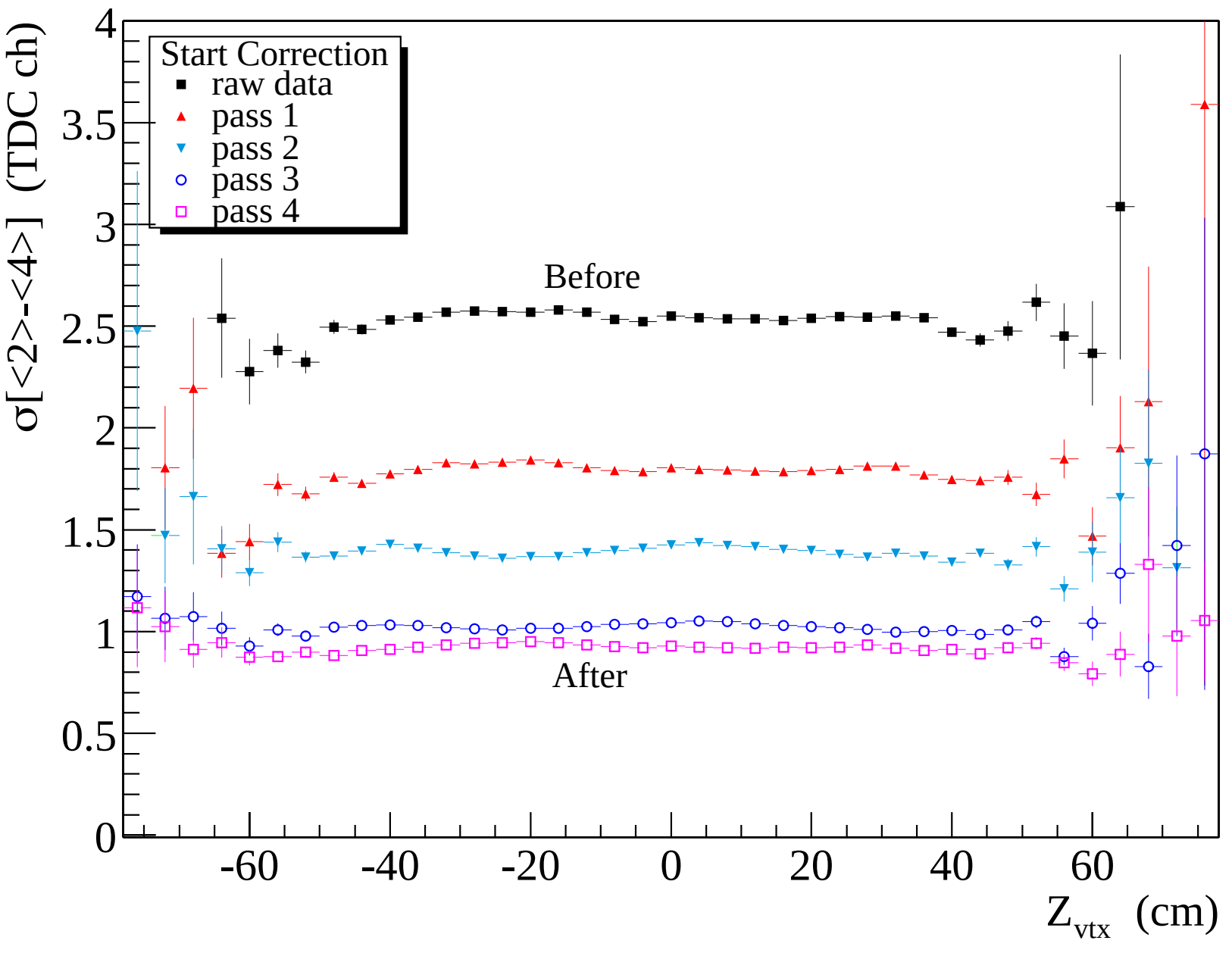

Fig. 15. The standard deviation of the quantity ⟨2⟩−⟨4⟩ in units of TDC channels (50ps/ch) versus the primary vertex location. The different point styles correspond to individual passes during the start correction procedure.

is reduced to approximately 0.9 TDC channels, or ∼45 ps.

The start time needed by the TOFp system is the average of all six (slew-corrected) TDCs, *i.e.* "⟨6⟩". Assuming that all six pVPD PMTs have the same single-detector resolution, the standard deviation of the difference of averages ⟨2⟩−⟨4⟩ is related to the standard deviation of the average of all six pVPD PMTs by a factor of $\sqrt{4/3}/\sqrt{6}$, or 0.47. The pVPD start time resolution in the central full-energy Au+Au events was thus $\sigma_{\langle 6\rangle}$=24 ps, and the equivalent single detector resolution was $\sigma_{\langle 1\rangle}$=58 ps. This start-timing performance is considerably better than necessary to do efficient time of flight analyses in the STAR geometry. The corrections one then needs on the stop side are now described.

## *3.2 TOFp*

The TOFp slats also slew and are "single-ended." Thus required are corrections for the stop slewing and the the time taken for the scintillation light to propagate from the location of the hit inside the slat to the PMT. This distance is referred to as $Z_{hit}$ and it is inferred from the tracking information and the TOFp geometry. This propagation time was measured on the bench using a Nitrogen laser to be linear with a slope of ∼65 ps/cm. In the case of the STAR detector, the TOFp geometry, and RHIC events, however, these two corrections are correlated to some degree with a large number of other variables - requiring some care in the software used to correct for these effects. The TDC value depends on the actual time of flight of the particle in STAR, the scintillation light propagation time, and the shift from the slewing. The ADC value in a TOFp slat, which is assumed to correlate directly to the pulse height and hence to the slewing effect, also depends on a number of other quantities. It depends on the momentum via Bethe-Bloch-like energy loss. It depends on the location of the hit along the slat via the scintillation photon attenuation inside the wrapped scintillator. It depends on the length of the track inside the volume of the slat. This length also depends on the location of the slat in the TOFp tray (due to aspects of the angles of the slat positioning), the primary vertex location, and the track momentum components $p_T$ (in the bend plane) and $p_Z$ (in the non-bend-plane). Thus, in the case of TOFp in STAR, the slewing does not vary simply as const/$\sqrt{\text{ADC}}$ or const/ADC, and the scintillation light propagation time does not vary simply as const*$Z_{hit}$.

Given the numerous correlations between the TDC values and the ADC and $Z_{hit}$ values beyond those simply related to the slewing effect and the scintillation photon propagation time, we therefore use an iterative technique here as well. Polynomials in $\sqrt{\text{ADC}}$ and $Z_{hit}$ are fit to the data in several steps. The quantity that is minimized in this case is the difference between the experimental and expected $1/\beta = c\tau/s$ values for pions, where $c$ is the speed of light and $\tau$ is the measured TDC value minus an offset multiplied by the time-to-digital calibration of ∼50 ps/ch. The quantity $s$ is the primary track total path length calculated using the reconstructed track helix parameters and the STAR and TOFp geometry. All calibrations are defined by optimizing the pions, and these corrections are then applied to all tracks in subsequent passes through the data. Pions for the calibration sample are selected by cuts on the difference of the experimental and expected $1/\beta$ values. The TPC dE/dx values are then required to be consistent w/ those for minimum ionizing particles at the same momentum.

The iterative technique provides a complete correction for the correlations between the average inverse velocity values and values of the variables ADC and $Z_{hit}$. The TDC channel-to-time calibration is performed using standard techniques based only on the experimental data itself. Temperature and drift corrections were explored but were not necessary for these data.

A plot of corrected track inverse velocity versus the track momentum following all these corrections is shown in Figure 16. The solid lines are the values expected based on the track momentum and the assumption of a mass - pion (lowest line), Kaon, proton, and deuteron (highest line). Strong bands for experimentally identified pions, kaons, and protons are seen with the expected resolution. The insets depict the projection onto the $1/\beta$ axis for specific momentum regions that are near the limits of TOFp's reach. Well separated peaks for the various hadrons are apparent.

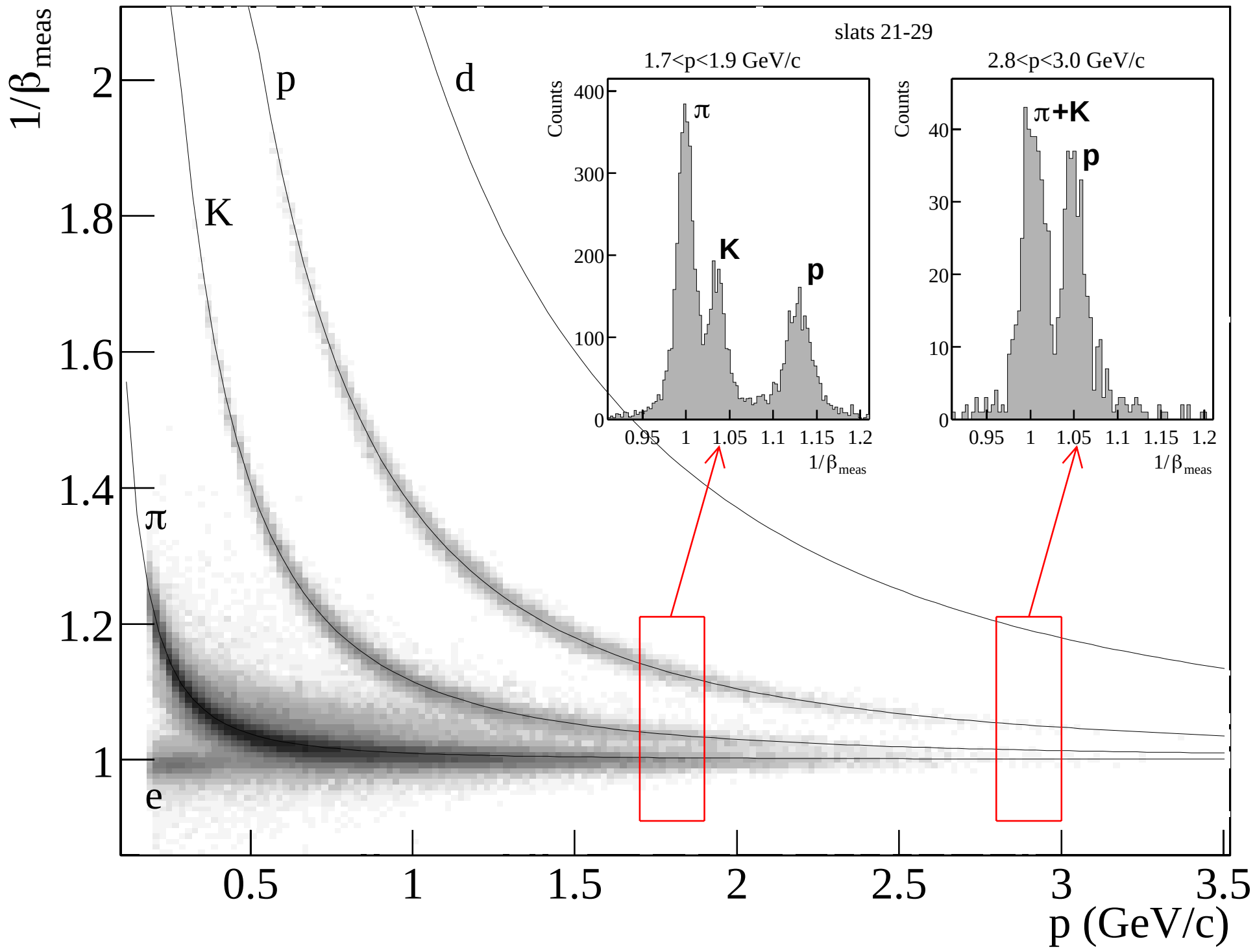

Fig. 16. The inverse velocity versus momentum for one TOFp slat following the start and stop corrections. The insets depict the projection onto the $1/\beta$ axis for specific momentum regions.

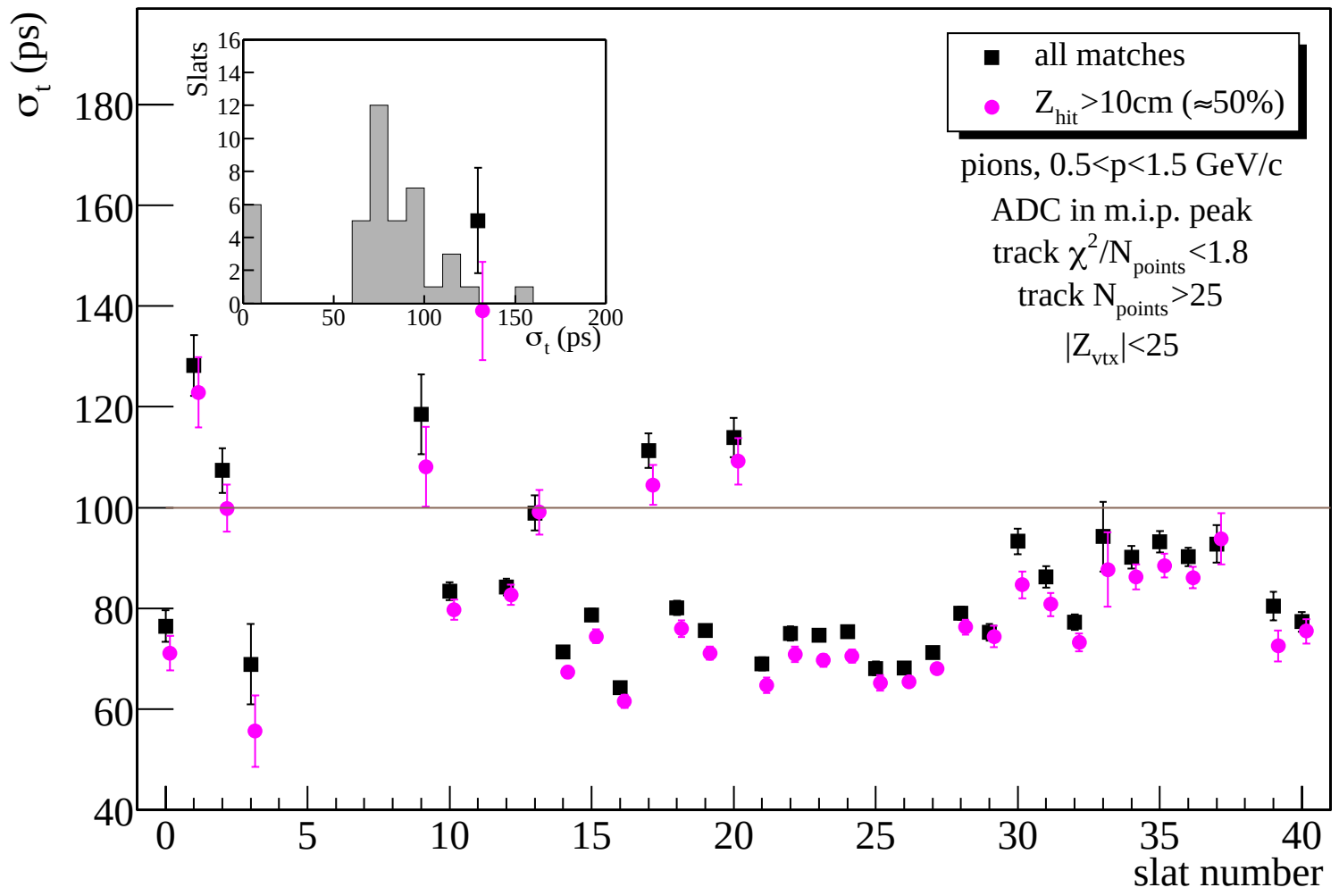

Fig. 17. The total time resolution of the system versus the slat number. The inset histogram depicts the number of slats versus the time resolution.

Shown in Figure 17 is the total time resolution for 0.5-1.5 GeV/c pions versus the slat number after all corrections. The slats with numbers 0-4 are in the first row near $\eta \sim 0$, while the slats with numbers $37-40$ are in the last row near $\eta \sim 1$. The squares depict the resolution for all tracks that pass the track quality cuts listed in the figure. There is a slightly better time resolution for those matches that are not close to the PMT in a given slat - the circles depict the resolution with an additional cut that $Z_{hit}$ is greater than 10 cm.

There are a few poorly-performing channels and one FEE board that was off for this run (slats 5-8). All of these have since been repaired. The rest of the slats have a total time resolution near or below 100ps. The average time resolution of the system over all slats is 87 ps. The average time resolution over the best 25 slats is 79 ps.

## 4 Summary

A small acceptance Time-Of-Flight system was added to the STAR experiment at RHIC for the measurement of charged hadron rates in RHIC Au+Au collisions with direct particle identification. In certain ranges of the particle momentum, *i.e.* 0.6-1.0 GeV/c for Kaons and 1.0-1.5 GeV/c for protons, the present system fills gaps in STAR's PID capabilities between dE/dx PID in the TPC and Čerenkov-ring PID in the RICH. Above and below these gap regions the PID capabilities of this TOF system are complementary to the other techniques, allowing cross-checks between the various PID approaches in STAR.

Both start (pVPD) and stop (TOFp) detectors were added to STAR. The detector design in both subsystems involves the conventional design of fast plastic scintillator and PMTs. Custom high-performance FEE were developed for this system which outperform the common commercial discriminators. The signal path involves Belden 9310 cable, which was shown to outperform other common coaxial cable types, and re-discrimination along the signal path, in order to optimize the timing performance at the digitizers. A local trigger system implemented in NIM logic, and a local data acquisition system based on CAMAC ADCs and TDCs and custom Linux-based software, was developed to provide the necessary interfaces to the STAR Trigger and DAQ systems.

The design goal of the system was to achieve a total time resolution near or better than 100 ps, which in the STAR experiment would allow direct $\pi$ *vs.* K *vs.* p identification for momenta from ~0.3 GeV/c to ~1.7(1.9) GeV/c, and direct p *vs.* ($\pi$+K) identification up to ~2.6(3.1) GeV/c, for tracks near pseudorapidity $\eta \sim 0(1)$. In full energy central Au+Au collisions, the start time resolution from the pVPD was observed to be ~24 ps, while the total resolution of the system TOFp (start and stop) was ~87 ps.

## Acknowledgements

We thank the STAR Collaboration for the use of the experimental data from the RHIC 2001 run shown in section 3, and for the operation of the TOFp & pVPD Systems as part of STAR standard shift crew operations. We are grateful for the technical assistance of members of the STAR Technical Support Group including K. Asselta, R. Brown, B. Christie, T. Krupien, A. Lebedev, D. Padrazo, J. Scheblein, and R. Soja. We are also grateful for the expert assistance and helpful comments from members of the STAR collaboration including T. Hallman, H. Crawford, E. Judd, F. Bieser, A. Ljubičič, Jr., P. Jacobs, T. Trainor, B. Lasiuk and J. Dunlop. We thank V. Astakhov (JINR-Dubna) for consistent support regarding the HVSys high voltage system. We thank E.S. Smith (TJNAF) for helpful comments on the coaxial cable simulations. We thank P. Reynolds and P. Novakova for administrative assistance throughout the project. We gratefully acknowledge funding from Brookhaven National Laboratory under Grant Number 12132, from the US Department of Energy under Grant numbers DE-FG03-96ER40772 and DE-FG02-89ER40531, and from the National Science Foundation of China under Grant numbers 19775018 and 10275027.

## References

[1] K.H. Ackermann *et al.*, Nucl. Instrum. Meth. A **499**, 624 (2003);
J. Harris *et al.*, Nucl. Phys. **A698**, 64 (2002).

[2] H. Wieman *et al.*, IEEE Trans. Nucl. Sci. **NS-44** 3, 671 (1997);
M. Anderson *et al.*, Nucl. Instrum. Meth. A **499**, 659 (2003).

[3] F. Bergsma *et al.*, Nucl. Instrum. Meth. A **499**, 633 (2003).

[4] A. Braem *et al.*, Nucl. Instrum. Meth. A **499**, 720 (2003);
J. Dunlop *et al.*, Nucl. Phys. **A698**, 515 (2002).

[5] S. Ahmad *et al.*, Nucl. Inst. Methods A **330**, 416 (1993).

[6] V. Astakhov *et al.*, http://www.hvsys.dubna.ru/.

[7] S.P. Stoll, PHENIX Internal Note No. 245, July 1, 1996;
S.P. Stoll, PHENIX Internal Note No. 245A, April 8, 1997.

[8] Last-A-Foam FR-6700, 3 lbs/cu.ft. (General Plastics Company, Tacoma, WA 98409).

[9] B. Bengtson and M. Moszynski, Nucl. Inst. Methods **81**, 109 (1970);
M. Moszynski and B. Bengtson, Nucl. Inst. Methods **158**, 1 (1979).

[10] T. Sugitate *et al.*, Nucl. Inst. Methods A **249**, 354 (1986).

[11] R.E. Bell, Nucl. Inst. Methods A **42**, 211 (1966);
M.A. El-Wahab *et al.*, Nucl. Inst. Methods A **78**, 325 (1970).

[12] S. Albergo *et al.*, Nucl. Inst. Methods A **362**, 423 (1995).

[13] R. Lacasse *et al.*, Nucl. Inst. Methods A **408**, 408 (1998).

[14] K. Shigaki *et al.*, Nucl. Inst. Methods A **438**, 282 (1999).

[15] E.S. Smith *et al.*, Nucl. Inst. Methods A **432**, 265 (1999).

[16] http://www.phenix.bnl.gov/phenix/WWW/tof/index.html;
K. Ikematsu *et al.*, Nucl. Inst. Methods A **411**, 238 (1998).

[17] G. Fidecaro, Nuov. Cimento 15 Series X (Suppl.), 254 (1960).

[18] E.S. Smith, CEBAF Technical Note TN-91-022, April 1991.

[19] CERN Program Library, KERNLIB routine RFFT (D703).

[20] In Ref. [18], the constant $c_0$ is also identified as $R_{dc}/2Z$, where $R_{dc}$ is the direct current resistance of the center conductor and Z is the impedance.

[21] F.S. Bieser *et al.*, Nucl. Instrum. Meth. A **499**, 766 (2003).

[22] J.M. Landgraf *et al.*, Nucl. Instrum. Meth. A **499**, 762 (2003).

[23] W.J. Llope *et al.*, Nucl. Inst. Methods A **443**, 451 (2000).

[24] R. Brun *et al.*, http://root.cern.ch/.